# Does the Outer Region of the Turbulent Boundary Layer Display Similar Behavior?

David Weyburne
AFRL/RYDH
2241 Avionics Circle
WPAFB, OH 45433

**ABSTRACT**

Recent theoretical results together with established theory have identified the displacement thickness and the velocity at the boundary layer edge as similarity scaling parameter candidates for the wall-bounded turbulent boundary layer. In the work described herein, we examine these scaling parameters along with the Prandtl Plus scaling's and the Zagarola and Smits scaling's to search for similarity in the outer region of experimental turbulent boundary layer velocity profile datasets. A new integral area method combined with the traditional chi-by-eye method is used to search for similar velocity profiles. The results indicate that strict whole profile similarity is not evident in any of the datasets we searched. However, ten datasets are found that display "similar-like" behavior using the ratio of the inner to outer thickness ratio as a search criterion. In alignment with theory, the preferred similarity scaling parameters for the similar-like behavior case are the displacement thickness and the velocity at the boundary layer edge. It was found that there are a few datasets for which the Prandtl Plus scaling and the Zagarola and Smits scaling also work.

## 1. INTRODUCTION

One of the most fundamental concepts in fluid mechanics is to analyze experiment observables using dimensional analysis with the intent of finding scaling parameters such that the scaled observable from different stations along the flow appear to be similar. Similarity of the velocity profile formed by fluid flow along a wall is one of those fundamental observables. For 2-D wall-bounded flows, velocity profile similarity is defined as the case where two velocity profiles taken at different stations along the flow differ only by simple scaling parameters in $y$ and $u(x,y)$, where $y$ is the normal direction to the wall, $x$ is the flow direction, and $u(x,y)$ is the velocity parallel to the wall in the flow direction. Following Schlichting [1], a velocity profile at position $x_1$ is said to be similar to the velocity profile at $x_2$ if

$$\frac{u(x_1, y/\delta_s(x_1))}{u_s(x_1)} = \frac{u(x_2, y/\delta_s(x_2))}{u_s(x_2)} \quad \text{for all y,} \tag{1}$$

where the length scaling parameter is $\delta_s(x)$, and the velocity scaling parameter is $u_s(x)$.

There has been a considerable effort in the literature to discover the $\delta_s(x)$ and $u_s(x)$ parameters for different wall-bounded flow situations. Theoretical similarity solutions of the flow governing equations are well known for laminar flow. Turbulent flow similarity is more problematic. One of the complicating factors for the turbulent boundary layer is that it appears to be composed of two distinct regions, one in the near wall region where viscosity effects are important and an outer region where they are not important. It is generally accepted that the wall-bounded turbulent boundary layer velocity profiles do not show whole profile similarity due to the viscosity effects in the inner region area [2]. As a result, it has been common to consider similarity issues for the inner and outer region separately [3]. The recent work of Weyburne [4] indicates that most experimental wall-bounded turbulent boundary layer datasets do NOT show similarity in the outer region. Hence, there are a number of questions that need to be answered; are there any turbulent boundary layer datasets show similarity in the outer region, how does one identify likely datasets, and what are the appropriate similarity scaling parameters?

[1]Email: David.Weyburne@gmail.com

The experimental search for similarity is hampered by the fact that it is very difficult to use Eq. 1 directly since the "stretched" *y*-values do not, in general, match from station to station. The problem is that while one measures the velocity at fixed heights *y* above the wall in the wind tunnel, the stretched value $y/\delta_s(x)$ will not, in general, correspond to the stretched value at another location since $\delta_s(x)$ is varying along the wall. Hence, Eq. 1 is not usable directly in most cases for experimental comparison purposes.

Because Eq. 1 is not usable for searching for similarity in experimental turbulent boundary layer datasets, an alternative method involving plotting all of the scaled profiles on one graph has become common in the literature. What is done is to view the plotted scaled velocity profiles in order to see whether the profiles plot on top of one another. A set of scaled profiles that display similar behavior should appear to be nearly identical. This check for similarity is what is sometimes referred to as the visual "chi-by-eye" method, *i.e.* do the plots overlap convincingly by visual observation. Of course, this makes the chi-by-eye method subjective and readily influenced by the method of plotting and the plotting scales. For example, in Fig. 1a we reproduce a figure from the literature that the authors [5] contend shows similar behavior of a set of seven scaled experimental turbulent velocity profiles when plotted using the Prandtl Plus scaling ($y^+ = yu_\tau/\nu$, $u^+ = u/u_\tau$, where $\nu$ is the kinematic viscosity and $u_\tau$ is the friction velocity). Looking at Fig. 1a one may, in fact, agree with the author's assessment that similarity is indeed present. However, if we take the same data and use the same scaling parameters but replot the data using a linear instead of Log scale and use plotted lines instead of symbols, one obtains Fig. 1b. Our chi-by-eye subjective assessment of Fig. 1b is that similarity is **not** present in this dataset.

In the work herein, we propose a different, but complimentary, approach for finding similarity in turbulent velocity profile datasets that removes much of the subjectivity present in the chi-by-eye plotting method. The new approach is based on an integral moment method to study similarity recently developed by Weyburne [6]. The approach is based on a simple concept, the area under a set of scaled profile curves that show similar behavior must be equal. By taking various integrals of the similarity defining equation (Eq. 1), it is possible to find new properties of the scaled profiles which must be true if similarity is present in a set of velocity profiles. It was demonstrated, for example, that for any 2-D flow displaying similarity, the length scaling parameter $\delta_s(x)$ must be proportional to the velocity displacement thickness, $\delta_1(x)$, and the velocity scaling parameter $u_s(x)$ must be proportional to the velocity $u_e(x)$, the velocity at the boundary layer edge [6]. Herein, we propose to use the area under the scaled profiles as a simple similarity test. The area under each velocity profile is first calculated and then the coefficient of variation, CV, (the standard deviation divided by the mean), is calculated for each candidate dataset. Similarity is indicated when the CV is smaller than a threshold that is chosen by examination of profile plots.

Some would argue that the search for similarity for the turbulent boundary layer has already been completed. A number of recent turbulent scaling reviews/papers [7-10] indicate that Zagarola and Smits [11] scaling parameter seems to work on almost all turbulent boundary layer datasets. However, recent revelations by Weyburne [4,12,13] cast serious doubt on the Zagarola and Smits scaling velocity parameter as used in the literature. The problem, as outlined by Weyburne, is that the Zagarola and Smits velocity scaling parameter, $u_{ZS}(x)$, does not always satisfy all of the defect profile similarity requirements. What has been missing from the literature is a realization that every defect profile based theoretical paper on turbulent boundary layer similarity [2,6,7,14] has indicated that similarity requires that any candidate scaling parameter, such as $u_{ZS}$, must be proportional to $u_e(x)$ at each measurement station. Unfortunately, for whatever reason, this similarity requirement has been ignored. When Weyburne [4,12] did cross-check many similarity claims from the literature, it was found that this condition was NOT satisfied. The problem has gone unnoticed because the defect profile plots hide the fact that $u_e(x)/u_{ZS}(x)$ is not equivalent from station to station. In the tail region of the scaled profile plots, you will see the $u_e(x)/u_{ZS}(x)$ ratio value directly whereas for the tail region of the defect profile, the tail is being intentionally forced to zero. Hence any differences in the defect profile tail region are too small to notice. Consequently, the defect profile plots look very good even if the $u_e(x)/u_{ZS}(x)$ values are not equal. If $u_e(x)/u_{ZS}(x)$ values are not equal then similarity is NOT present [2,6,7,14] no matter how good the defect profile plots look. In the work herein, we explicitly check $u_{ZS}$ along with $u_e$ and the friction velocity $u_\tau$ to determine which velocity scaling parameters do work for the turbulent boundary layer outer region.

A second finding from Weyburne's [4,12,13] work on similarity scaling parameters is also related to the fact that any candidate scaling parameter, $u_s$, is theoretically required to satisfy the condition that the $u_e/u_s$ values must be equal from station to station. This requirement necessarily implies that it is not possible to have defect profile similarity unless velocity profile similarity is already present [4,12]. It is a direct consequence of the definition of defect similarity given by

$$\frac{u_e(x_1)-u\left(x_1, y/\delta_s(x_1)\right)}{u_s(x_1)} = \frac{u_e(x_2)-u\left(x_2, y/\delta_s(x_2)\right)}{u_s(x_2)} \quad \text{for all y,} \tag{2}$$

together with the fact that the $u_e/u_s$ values must be equal. By inspection of Eqs. 1 and 2, it is apparent that defect profile similarity necessarily requires that velocity profile similarity must also be present. This is an astounding revelation given that the fluid flow community has exclusively pursued the use of the defect profile for studying turbulent boundary layer similarity for more than a half a century to the extent that to even discuss velocity profile similarity was considered wrong and showed one's lack of understanding of turbulent boundary layer theory. Weyburne discusses how this omission may have persisted for so long [13] but for the purpose herein, the message is clear; similarity of the turbulent boundary layer is properly studied using the experimentally measured velocity profile.

In what follows, we start by developing the equal area method's similarity equations. We then outline the new statistical method for searching for similarity in turbulent boundary layer datasets. A number of experimental datasets are then examined using this statistical approach.

## 2. Similarity of the Velocity Profile

The intent of Schlichting's [1] definition of similarity is clear but it invokes an unspecified redefinition of the velocity. For our purposes, a clearer definition is instructive: the velocity profiles at positions $x_i$ and $x_j$ are said to be similar if there exist certain scaling parameters $\delta_s(x)$ and $u_s(x)$ such that

$$\frac{\bar{u}(x_i, y_i)}{u_s(x_i)} = \frac{\bar{u}(x_j, y_j)}{u_s(x_j)} \quad \text{where} \quad y_i = y_j, \quad \bar{u}(x_k, y_k) = u(x_k, y), \quad \text{and} \quad y_k = \frac{y}{\delta_s(x_k)} \quad \text{for all } y. \tag{3}$$

This definition emphasizes the fact that we are comparing the scaled velocity values at equivalent $y_i$-values and not equivalent *y*-values. With this definition, it is self-evident that if the profiles are similar, then the area under the scaled velocity profiles plotted against the scaled $y_i$-values must be equal. If we multiply both sides of Eq. 3 by $y_i$ raised to the *n*-th power, then the two sides are still equivalent. Hence, the area under the scaled profiles times the scaled $y_i$-value to the *n*-th power must also be equal if similarity is present. The area under the scaled velocity profile times the scaled $y_i$-value to the *n*-th power at $y_i$, in integral form, is given by

$$a_n(x_i) = \int_0^h dy_i \; y_i^n \frac{\bar{u}(x_i, y_i)}{u_s(x_i)} \quad , \tag{4}$$

As long as the limit value, *h*, are equal and are all deep into the free stream above the wall then a necessary **and** sufficient condition for similarity [6] is that

$$a_n(x_i) = a_n(x_j) \qquad n = 0,1,2,...,\infty \quad . \tag{5}$$

To definitively prove similarity, one must calculate $a_n(x)$ for all *n*. However, to merely find likely datasets displaying similarity, it is only necessary to calculate and compare $a_n(x)$ at a single *n*-value. For the *n*=0 case, it is easy to show mathematically that for the combination $\delta_s(x) = \delta_1(x), u_s(x) = u_e(x)$, the calculated values for the $a_0(x)$ are automatically equivalent from station to station. Hence, the *n*=0 case is not ideal to search for similarity in experimental datasets. Therefore, we will concentrate on our similarity search using the *n*=1 case using Eq. 4.

The procedure for searching for similarity is to calculate the $a_1(x)$ integral for each profile in a dataset. From there, the mean value and the standard deviation are calculated for each group of profiles. The coefficient of variation (CV) is then calculated. The CV is used as our figure of merit. A CV value near zero would imply the dataset displays similarity for that $\delta_s(x)$, $u_s(x)$ combination.

### 3.1 Finding potential datasets

The first step in our search for similarity is to identify likely datasets. The search for similar datasets could have proceeded as Clauser [15] and Castillo and George [7] did by looking for datasets which satisfy the appropriate pressure gradient condition. Of course, that strategy does not work for the zero-pressure gradient (ZPG) cases. Rather than excluding the ZPG cases, we prefer the simpler approach of looking for datasets that display a region where the inner and outer boundary layer regions change proportionately from station to station. In the past, this inner/outer ratio test would have been based on checking whether the Rotta [14] ratio $u_e(x)/u_\tau(x)$ is constant. The Rotta ratio assumes the thicknesses of the inner and outer region are proportional to the characteristic velocities of the two regions. However, it has never been proven that having a constant Rotta ratio is a similarity requirement.

As an alternative, a new method for describing the boundary thickness and shape [16] will be used. This new method provides an experimentally accessible measurement of this thickness ratio. For the new thickness ratio, we use $\delta_v = \mu_1 + 2\sigma_v$ as the thickness for the viscous inner region and $\delta_d = \delta_1 + 4\sigma_s$ as the thickness of the outer region [16]. The inner region $\mu_1$ and $\sigma_v$ values can be calculated in terms of $u_e$, $u_\tau$, and the kinematic viscosity [16]. The boundary layer width is calculated as $\sigma_s^2 = 2\alpha_1 - \delta_1^2$ where $\delta_1$ is the displacement thickness and where

$$\alpha_1(x) \;=\; = \int_0^h dy\; y\left\{1 - u(x,y)/u_e\right\} \quad . \tag{6}$$

This thickness ratio $\delta_v/\delta_d$ is therefore straightforward to calculate from known or easily calculated parameters. A big advantage of this new thickness ratio is that using the integral area method, it is possible to prove that similarity requires that $\delta_v/\delta_d$ must be constant if similarity is present in the set of velocity profiles [17]. The search for likely datasets consisted of looking for datasets where the thickness ratio $\delta_v/\delta_d$ was constant or almost constant as a function of the Reynolds number.

### 3.2 Numerical calculation of areas

Once a set of likely profiles was identified, it was then necessary to calculate $a_1(x)$ for each profile. It was found that for most experimental velocity profiles, the normalized values outside the boundary layer region (*i.e.* nominally in the free stream) often fluctuated about the $u_e(x)$ value. Our experience has shown that noise in the high *y*-range velocity profile data can lead to errors in the calculated areas. So, to avoid this problem, we rounded all the velocities that were nominally in the free stream to exactly $u_e(x)$. The free stream starting point was taken as the first point at which the velocity ratio $u/u_e$ was greater than or equal to one. The limit value *h* for the calculation in Eq. 4 was picked to ensure all of the profile limits were in the free stream for that *h*-value choice. This was done by first identifying the profile with the largest boundary layer thickness $\delta_d(x)$. The *h*-value for that whole group was then assigned as the first $y/\delta_s(x_i)$ value for which $u/u_e$ was greater than or equal to one for that profile. The data point $\bar{u}(x_i,h) = u_e(x_i)$ was added as needed to the whole group so that the upper integrand limits in Eq. 4 all matched. For the numerical calculation of the integrals reported herein, the Trapezoidal Rule was used and the data point, $u(x,0) = 0$, was added to every dataset.

## 4. Results

### 4.1 Dataset screening results

In looking through experimental datasets, we could not find any datasets for which $\delta_v / \delta_d$ or $u_e / u_\tau$ were truly constant with Reynolds number. This result reinforces earlier observations [2,3] that whole profile similarity does not seem possible for the wall-bounded turbulent boundary layer. What we did find is presented in Fig. 2 and Table 1. All the examined datasets showed a steady increase in the $\delta_v / \delta_d$ ratio's value as the Reynolds number increases (the same is true for $u_e / u_\tau$, but not shown). Therefore, we choose a slightly relaxed similarity criterion that requires the CV to change by no more than about $\pm 10\%$ from its mean value. A total of ten datasets or parts of datasets were considered in the work herein which met this condition. Four other datasets were included because other groups have identified the dataset as possessing outer region similarity.

### 4.2 Similarity results

The results of the calculations for $a_1(x)$ coefficient of variation (CV) values for fourteen datasets are given in Table 2. The first ten datasets (names in blue) meet our dataset selection criterion. The datasets in Table 2 are ordered by their CV values for the $a_1(x)$ values for $u_e$ since theory indicates that this should be the correct scaling. For comparison, we also include plots of the datasets from Table 2 as Figs. 3-16. In the Introduction, we indicated the new similarity method removes much of the subjectivity inherent in similarity selection but it does not remove it all. The subjectivity problem encountered in the new method is where do we set the cut-off criteria. What CV value should indicate a similar dataset? In theory, if we had all of the experimental error information for all of the datasets then we could set our selection criteria for similarity based on the error bar information. Unfortunately, while a few datasets include the required information, not enough of them report experimental error bars to be able to use this method to set CV criteria. Hence, we are forced to make a subjective cutoff based on examination of the Figs. 3-16 together with an examination of Table 2. Although not definitive, **similarity** may be indicated for **$a_1(x)$ CV values less than 0.005**. To designate which of the $a_1(x)$ CV values meet that criterion, we marked those values in **red** in Table 2.

At the beginning of this research effort, we anticipated that the integral area method would be able to completely replace the chi-by-eye method. This turned out not to be the case. The mitigating problem of the area method is something we call the symmetry point problem. It is most easily seen in Figs. 13a-16a. If one looks at the section of curves to the left of the symmetry point, the curves all line up with the highest Reynolds number curve having the largest amplitude. To the right of the symmetry point, the order reverses and the lowest Reynolds number profiles have the largest amplitude. Technically, this symmetry point behavior means the datasets do not exhibit true similar behavior since they do not collapse to a single curve. The confounding area effect means that some datasets can show very low CV values, for example Fig. 9a, even though the curves do not completely overlap in a manner consistent with similar behavior. Although difficult to see at this scale, we did confirm that Fig. 9a, for example, does display the symmetry point behavior. It does not seem possible to use any of the calculated $a_n(x)$ integrals to detect this behavior. So, in the end, we must rely on visual confirmation on whether a dataset displays symmetry point behavior or not. To do this we examined blown-up sections of the curves (not shown) to see whether the plots displayed symmetry point behavior. Some of the datasets in red selected by the $a_1(x)$ CV values less than 0.005 did show symmetry point behavior (Figs. 9a and 11a). Those eight datasets marked in **red** and with an **asterisk** **did not show** blatant symmetry point behavior. These datasets passed all criteria we set and would appear to exhibit similar-like behavior.

## 5. Discussion

The results indicate that strict whole profile similarity is not evident in any of the datasets we searched. In every dataset we tested, the inner and outer regions were **not** changing proportionally along the wall as evidenced by the $\delta_v / \delta_d$ or $u_e / u_\tau$ values (Fig. 2 and Table 1). This was true even for the datasets where the authors [5,14] actively tried to impose similar behavior by adjusting the wind tunnel height along the flow direction. This evidence supports the general sense in the literature [2,3] that because the statistics of the turbulent structures are changing with Reynolds number for boundary layer flows, strict similarity in wall-bounded turbulent flow is not possible. So, it is probably the case that one can only hope to find flow conditions that closely approximate the required conditions (Fig. 2). An interesting example of this is seen in the Sillero, Jimenez, and Moser [18] DNS dataset. In Fig. 3a the data shows similar-like behavior that passes not only the chi-by-eye test but also has the smallest $a_1(x)$ CV value (Table 2) of all of the datasets we tested. Yet, if one rescales Fig. 3a using log-log scaling, Fig. 3d results. Although the differences are small, it is clear that the near-wall region judged from Fig. 3d shows non-similar behavior when scaled with $\delta_1$ and $u_e$. Hence, our conclusion is that while it is possible to find turbulent boundary layer datasets which come close to the right conditions and the datasets in fact show similar-like behavior in the outer region but in the end, they do not show true whole profile similarity.

For the last half century, the community has sought to establish the existence of scaling parameters $\delta_s(x)$ and $u_s(x)$ that collapse experimental turbulent velocity profile datasets to a single curve. The thinking was that since the equations for turbulent flows do not admit to exact similarity solutions, then one must search for parameters which collapse experimental data. However, very early on Rotta's [14] and Townsend's [2] theoretical treatments indicated that if similarity is present, then the velocity at the boundary layer edge $u_e$ must be a similarity scaling parameter. The subsequent theoretical development by Castillo and George [7] and Weyburne [3] support this result. The bottom line is that all the theoretical results for the similarity in 2-D turbulent boundary layer point to the velocity at the boundary layer edge $u_e$ as a similarity scaling parameter. These theoretical results support common sense: if the tail regions of the SCALED profiles from different stations do not have equal values then similarity is not present no matter what is happening in the rest of the profile. They can ONLY be equal if the velocity scale $u_s$ is proportional to $u_e$ since by definition $u_e$ is the value of the velocity profiles tail region. There have been challenges to the legitimacy of $u_e$ but those experimental challenges have been refuted [9,10]. What has not occurred is a challenge of the legitimacy of the supporting theories. Those who advocate for a velocity scaling parameter that is not $u_e$ must at some point explain where all the theories that have backed $u_e$ are wrong. Until the supporting theories have been proven wrong, it is only logical that we start our experimental search for turbulent boundary layer similarity by finding datasets for which the velocity scaling parameter is $u_e$. In a similar theoretical vein, Weyburne [3] showed that if velocity profile similarity is present in any wall-bounded 2-D flow, then the similarity thickness scaling parameter must be proportional to the displacement thickness $\delta_1$. The point to be made here is that contrary to popular belief, the turbulent boundary layer does have theoretical support for a set of scaling parameters for the wall-bounded turbulent boundary layer and those parameters are the displacement thickness $\delta_1$ and the velocity at the boundary layer edge $u_e$. It is therefore no coincidence that every experimental dataset that showed similar-like behavior showed this behavior predominantly for the $\delta_1$, $u_e$ combination as seen in Table 2. Although there were a few cases that showed similar behavior using the $u_{ZS}$ or $u_\tau$ based set of scaling parameters, all of those cases also showed similar behavior for the $\delta_1$, $u_e$ combination (Table 2). There were no examples found where similar behavior was **only** present for the $u_{ZS}$ or $u_\tau$ based set of scaling parameters (see red asterisk marked datasets in Table 2). Therefore, our experimental results support the theoretically based combination $\delta_1$, $u_e$ as similar-like scaling parameters.

It will be noticed that the similarity search was done on the velocity profile $u(x,y)$ rather than defect profile $u_e(x)-u(x,y)$. Let us be clear that this was not done in ignorance of the last half century of turbulent boundary layer similarity literature. We are fully cognizant of the flow community's belief that the important profile for the turbulent boundary layer is the defect profile and that to suggest using the velocity profile shows one's lack of understanding of turbulent boundary layer research. Incredibly, this stance, which developed over the last half century, is wrong. If one compares Eq. 2 to Eq. 1 then by inspection one sees the two definitions will be equivalent if $u_e/u_s$ is a constant as one moves from station to station. Indeed, every theoretical treatment for defect profile similarity has the requirement that $u_e/u_s$ must be constant [7, 11-13]. This necessarily means that one cannot have defect profile similarity unless velocity profile similarity is already present. Weyburne [9,10] discusses why this important point has been missed by the flow community but the theoretical requirements are clear; the important profile when it comes to similarity of the turbulent boundary layer is the experimentally measured velocity profile. One certainly can use the defect profile for similarity testing as long as one verifies that $u_e/u_s$ is constant.

There have been a number of groups that offered experimental evidence that the proper scaling for the outer region turbulent boundary layer is either $u_s=u_{ZS}$ or $u_s=u_\tau$. The results discussed herein in fact found that there are examples in the literature for which this scaling does appear to work. However, we would be remiss to not point out some of the literature claims that have been made erroneously. In the Introduction, we have already discussed the case made by Skåre and Krogstad [2] for the Prandtl plus scaling. Fig. 1b and its $a_1(x)$ CV value for $u_\tau$ (Table 2) makes it clear that this combination does not show similar-like behavior. A second example is the DeGraaff and Eaton [27] claim that their data supports the combination $u_s=u_{99}$ $u_s=u_\tau$ as similarity scaling parameters. Their evidence is a defect profile plot (their Fig. 4). Compare that to the velocity profile plot Fig. 15d below. It is evident that Fig. 15d does not show similar-like behavior. Looking at the tail region of Figs. 15a and 15d and the $u_e / u_\tau$ CV values (Table 1) indicates this ratio is not equivalent from station to station. Hence, the DeGraaff and Eaton [27] similarity claim is **not** supported for the combination $u_s = u_{99}$ and $u_s = u_\tau$. Next, consider the claim made by Cal and Castillo [9] that the FPG data of Schubauer and Klebanoff [23] shows similarity-like behavior for the combination $u_s = u_{99}$ and $u_s = u_{ZS}$. Contrast their Fig. 2b to Fig. 16b below. Fig. 16b does not show similar-like behavior. It is evident that the $u_e / u_{ZS}$ ratio is not constant and this result is confirmed by Table 1. The overall results indicate the FPG data of Schubauer and Klebanoff [23] does **not** show similar-like behavior.

Although we already discussed why we use the velocity profile instead of the defect profile, it is worth emphasizing this by looking at some of the plots. Consider some of the Super Pipe data by Zagarola, Smits, and co-workers [4,28]. Figs. 13b and Fig. 14b are plots of the velocity profiles taken in the Super Pipe facility at Princeton University's Gas Dynamics Laboratory using the Zagarola and Smits scaling $R$ (radius) and $u_{ZS}$ [4,28]. These plots do not show similar-like behavior. Now, if we change the plot scale to $u_{CL} - u_{ZS}$ ($u_{CL}$ is the center line velocity) instead of $u_{ZS}$, what happens is that near the center line all of those asymptotic near edge profile tails in Figs. 13b and Fig. 14b are forced to zero. What results is Figs. 13d and 14d. The defect profiles fall on top of one another, the chi-by-eye indication that similarity is present. However, the ratio $u_{CL} / u_{ZS}$, as evidenced in Fig. 13b and 14b (and Table 1), is not constant with Reynolds number. Hence these datasets do not pass the criteria for similarity even though the defect profiles look great. Our overall assessment of similarity claims made in the literature is that any claim made using the defect profile needs to be reexamined for the reasons discussed above.

The search criteria for similar datasets used herein is based on identifying conditions one would expect for whole profile similarity. Our conjecture is that similar-like behavior will only be observed for conditions where whole profile similarity conditions are almost met. We find that this "almost" condition is relatively rare. This is in contrast to the Castillo, George, and co-workers [7,8] paradigm where most turbulent boundary layer datasets display similarity behavior in the outer region. Which camp is correct? The consequences are far reaching in terms of computer modeling of fluid flow used to design and optimize aircraft, automobiles, etc. Direct numerical simulation of these complicated flow geometries will not be possible in the foreseeable future. Hence models of the boundary layer region are required to reduce computational effort to a manageable level. If the Castillo, George, and coworkers paradigm is correct then it might be possible to develop a robust boundary layer model for use in computer simulations. On the other hand, if turbulent boundary layer similarity is indeed rare, then this will not be possible.

## 6. Conclusions

No experimental datasets for wall-bounded turbulent boundary layer flows are found that exhibited strict similarity of the velocity profile. However, using an integral area method, datasets were found that show similar-like behavior in the outer region when the similarity criteria are closely, rather than exactly, met. In agreement with theory, the preferred scaling behavior appears to be the $\delta_1$, $u_e$ combination. There were a few examples found where $u_{ZS}$ or $u_\tau$ based scaling's also worked but there were no examples found where similar behavior was only present for the $u_{ZS}$ or $u_\tau$ cases.

**Acknowledgement**

The author acknowledges the support of the Air Force Research Laboratory and Gernot Pomrenke at AFOSR. In addition, the author thanks the various experimentalists for making their datasets available for inclusion in this manuscript.

# Tables and Figures

**Table 1: Coefficient of Variation (CV) for Selected Parameter Ratios**

| Dataset | CV $\delta_d/\delta_v$ | CV $u_e/u_\tau$ | CV $u_e/u_{ZS}$ |
|---|---|---|---|
| Sillero, Jimenez, & Moser [18] | 0.072 | 0.021 | 0.003 |
| Smith [21] | 0.115 | 0.030 | 0.031 |
| Jones, Marusic, & Perry [25] | 0.017 | 0.004 | 0.036 |
| Samuel & Joubert [22] | 0.059 | 0.054 | 0.031 |
| Zanoun & Durst [19] | 0.077 | 0.021 | 0.020 |
| Skåre & Krogstad [5] | 0.045 | 0.017 | 0.007 |
| Wieghardt & Tillmann [23] | 0.077 | 0.016 | 0.018 |
| Clauser [15] | 0.034 | 0.016 | 0.035 |
| Ludweig & Tillmann [26] | 0.071 | 0.019 | 0.043 |
| Herring & Norbury [20] | 0.039 | 0.008 | 0.031 |
| Zagarola [11] | 1.029 | 0.174 | 0.164 |
| DeGraaff & Eaton [28] | 0.597 | 0.113 | 0.098 |
| Hultmark, et. al. [29] | 0.645 | 0.116 | 0.136 |
| Schubauer & Klebanoff [24] | 0.326 | 0.368 | 0.244 |

**Table 2: Coefficient of Variation (CV) for $a_1(x)$ for $u_s$ Parameters**

| Dataset | CV for $u_e$ | CV for $u_\tau$ | CV for $u_{ZS}$ |
|---|---|---|---|
| Sillero, Jimenez, & Moser [18] | 1.6E-5* | 0.0211 | 0.0031* |
| Smith [21] | 0.0002* | 0.0233 | 0.0152 |
| Jones, Marusic, & Perry [25] | 0.0003* | 0.0028* | 0.0365 |
| Samuel & Joubert [22] | 0.0004* | 0.0448 | 0.1885 |
| Zanoun & Durst [19] | 0.0004* | 0.0123 | 0.0202 |
| Skåre & Krogstad [5] | 0.0008* | 0.0157 | 0.0100 |
| Wieghardt & Tillmann [23] | 0.0010 | 0.0097 | 0.0198 |
| Clauser [15] | 0.0011* | 0.0089 | 0.0382 |
| Ludweig & Tillmann [26] | 0.0012 | 0.0124 | 0.0788 |
| Herring & Norbury [20] | 0.0015* | 0.0043* | 0.0641 |
| Zagarola [11] | 0.0022 | 0.1777 | 0.1893 |
| DeGraaff & Eaton [28] | 0.0026 | 0.1051 | 0.0993 |
| Hultmark, et. al. [29] | 0.0047 | 0.1043 | 0.1410 |
| Schubauer & Klebanoff [24] | 0.0147 | 0.3695 | 0.2745 |

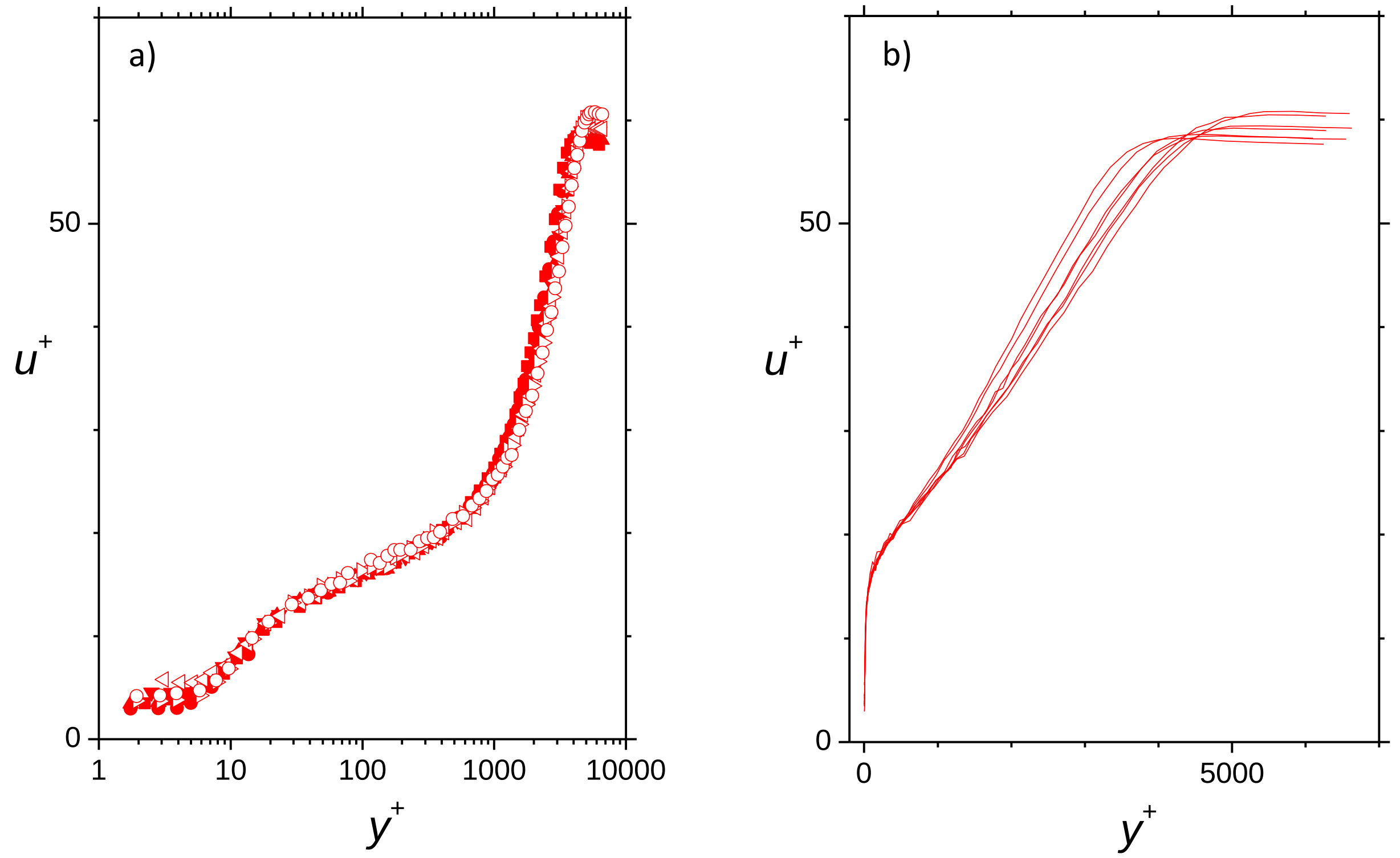

Fig. 1: Does this dataset show similarity? Skåre and Krogstad [5] say yes based on plot a). Plot b) shows the same data plotted using a linear *y*-scale and data points connected by lines.

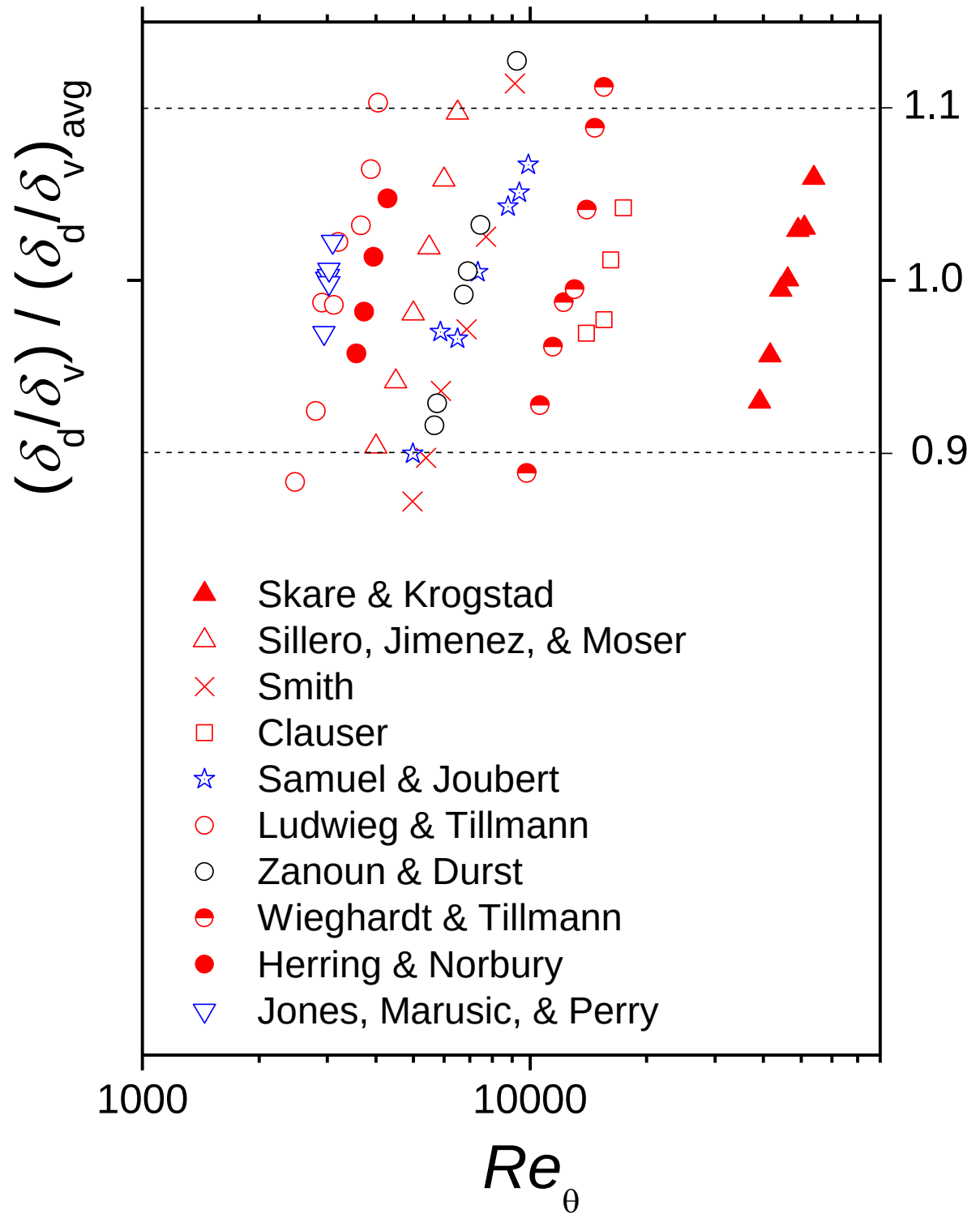

Fig. 2: Similarity criterion; normalized boundary layer thickness $\delta_d$ divided by the viscous boundary layer thickness $\delta_v$ for ten partial datasets as a function of Reynolds number.

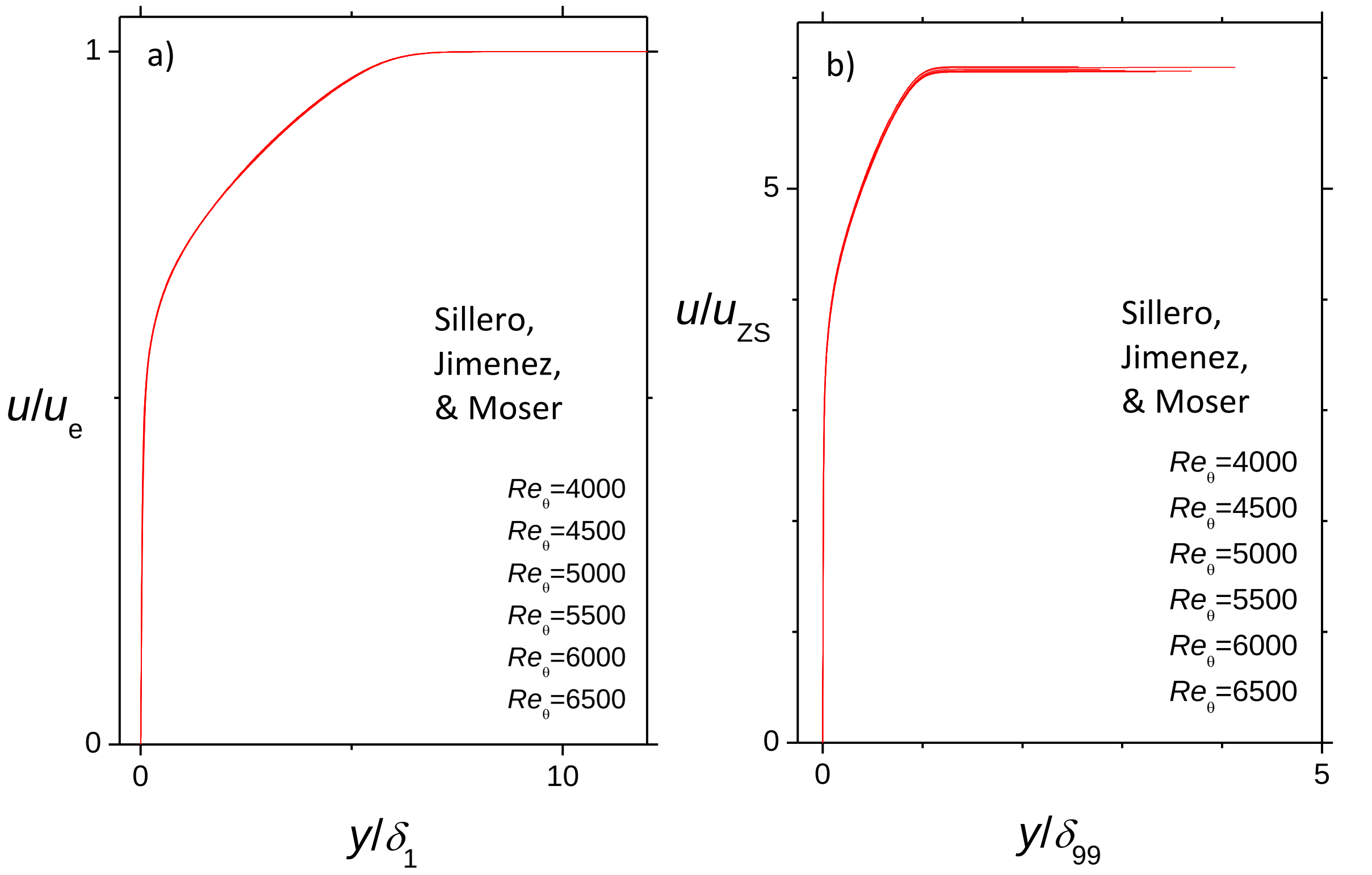

Fig. 3: Sillero, Jimenez, and Moser's [18] DNS data plotted a) using $u_s = u_e$, b) using $u_s = u_{ZS}$, c) using $u_s = u_\tau$, and d) using expanded $u_s = u_e$. The $Re_\theta$ identifies which profiles are plotted.

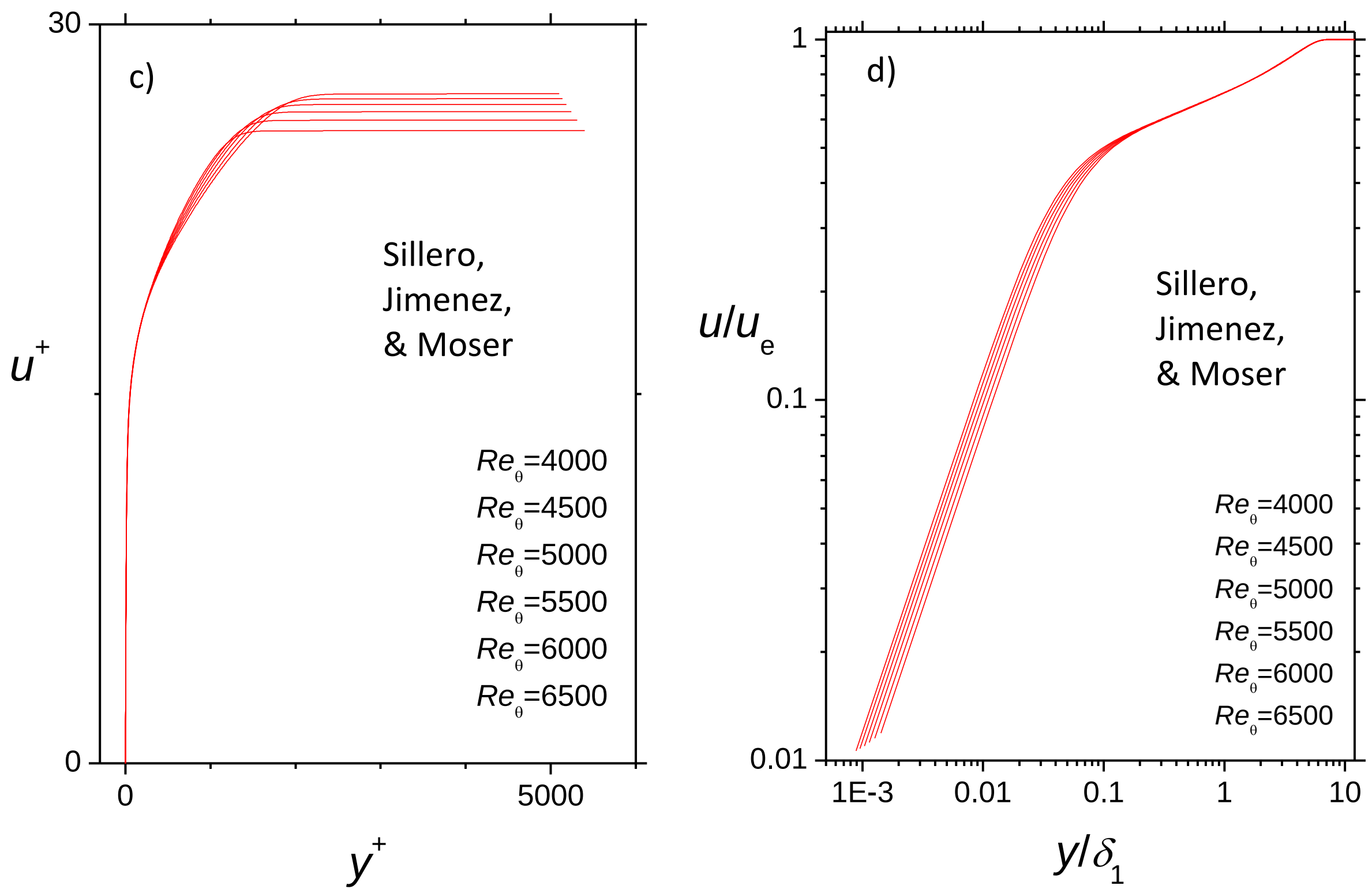

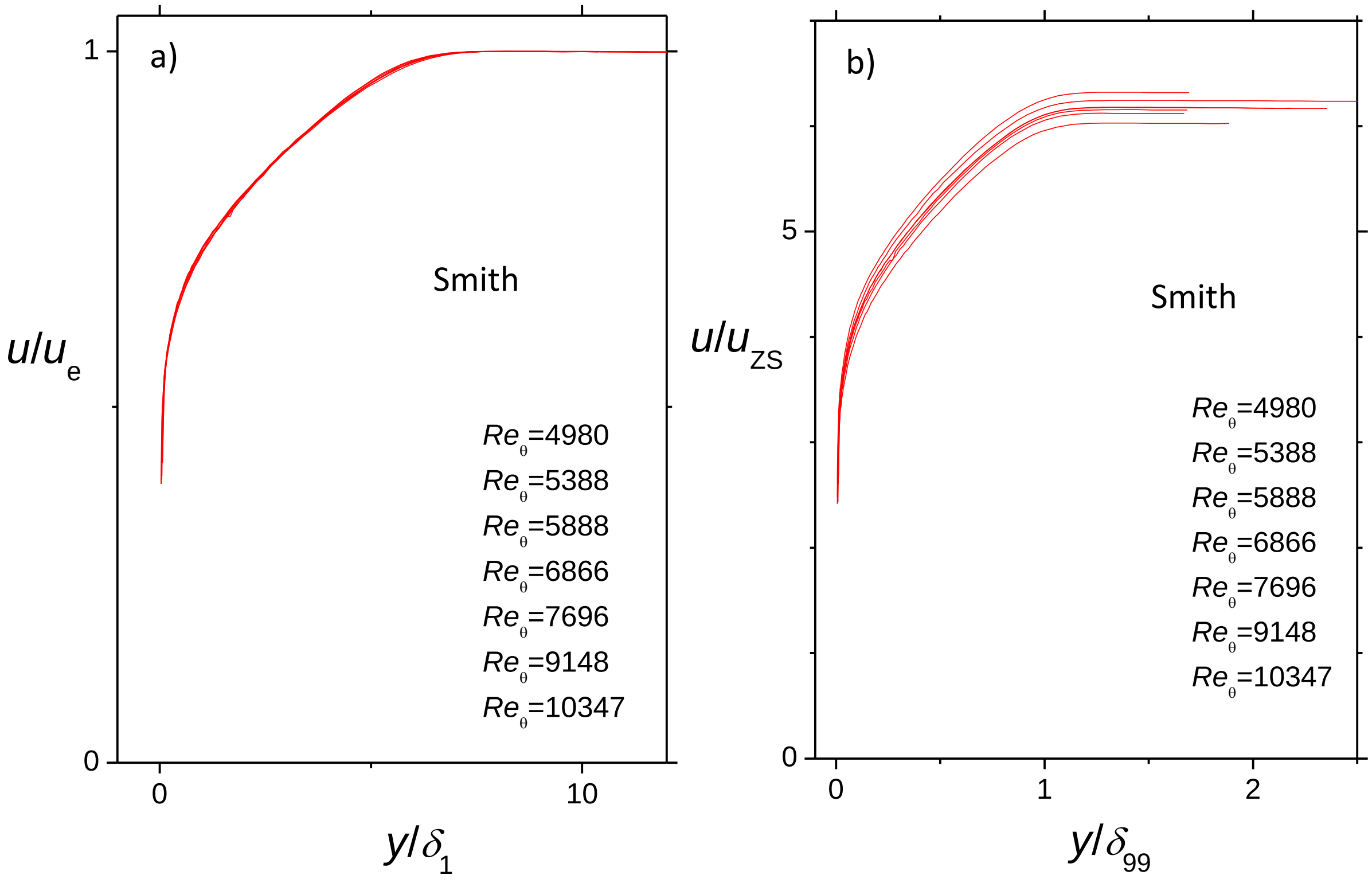

Fig. 4: Smith's [21] profile data plotted a) using $u_s = u_e$, b) using $u_s = u_{ZS}$ , and c) using $u_s = u_\tau$. The $Re_\theta$ list identifies which profiles are plotted.

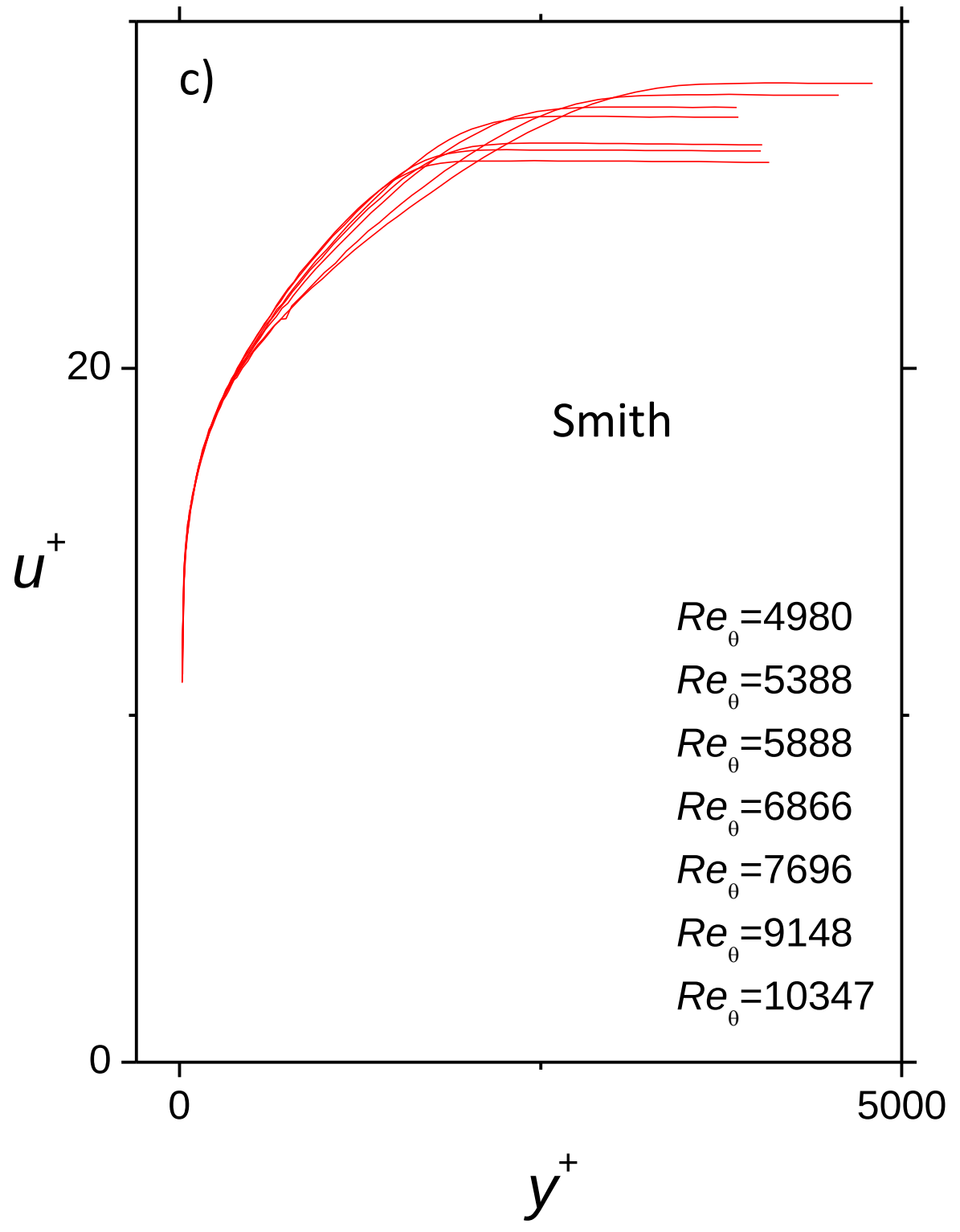

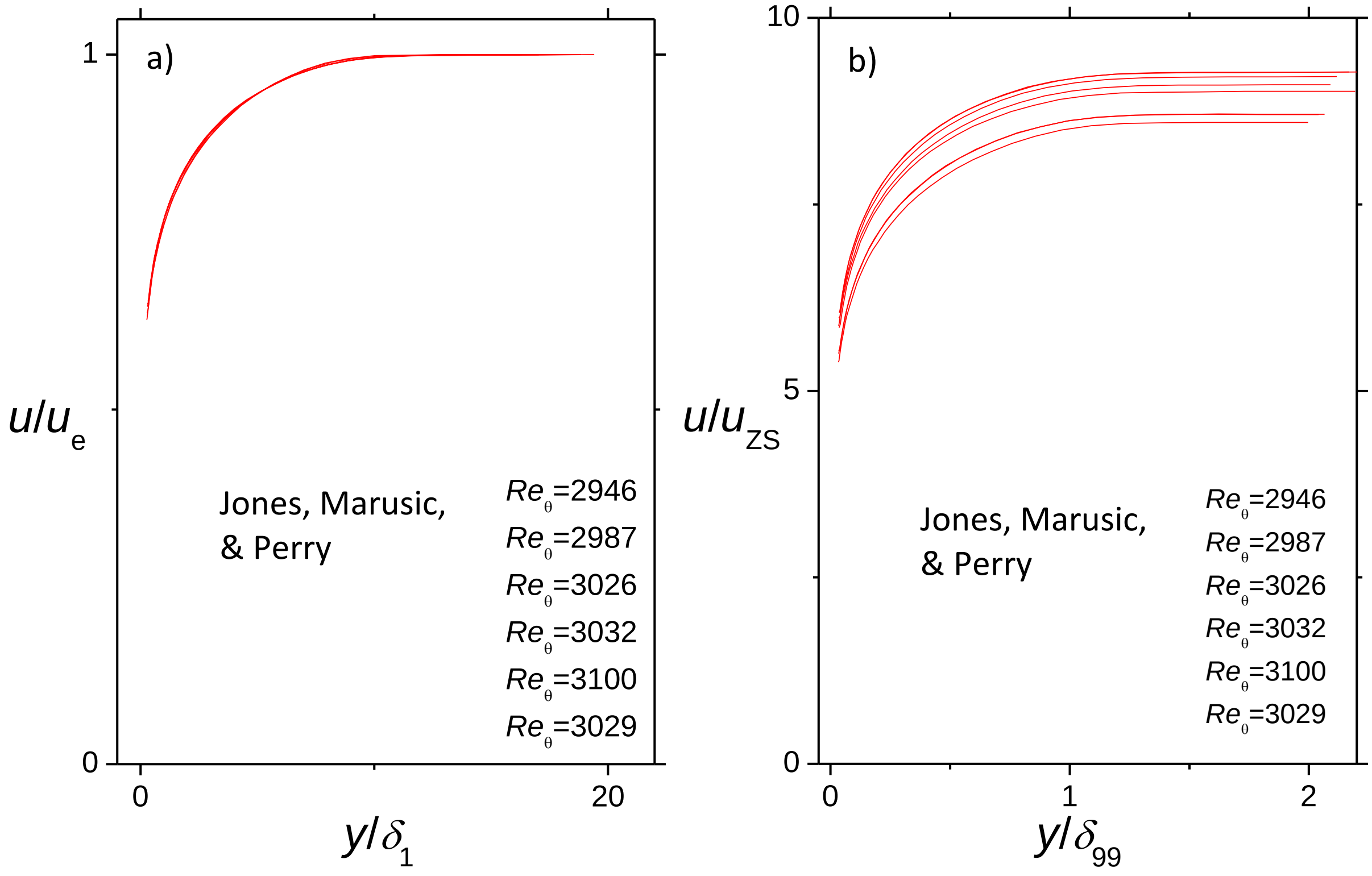

Fig. 5: Jones, Marusic, and Perry [25] profile data plotted a) using $u_s = u_e$, b) using $u_s = u_{ZS}$ , and c) using $u_s = u_\tau$. The $Re_\theta$ list identifies which profiles are plotted.

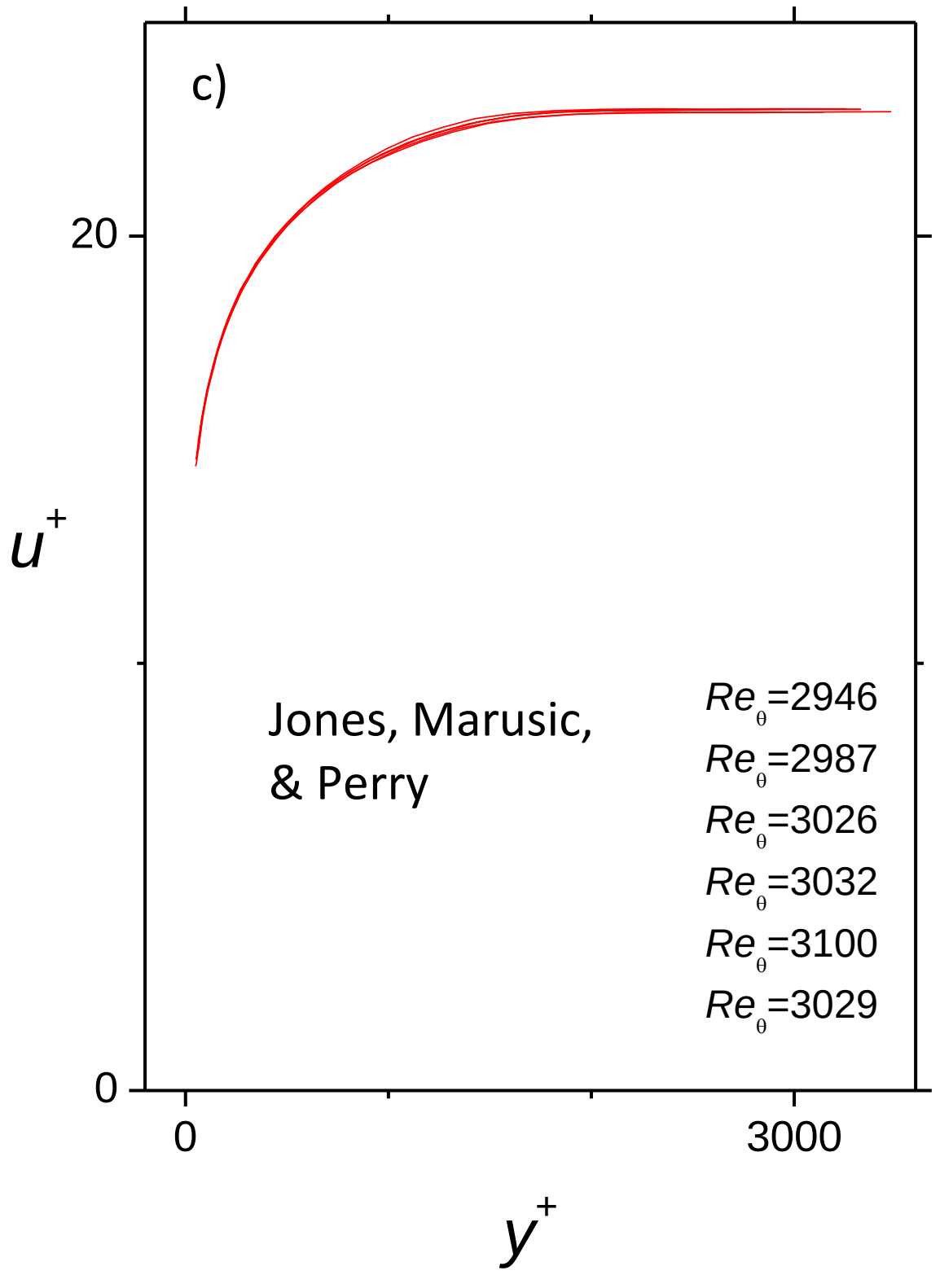

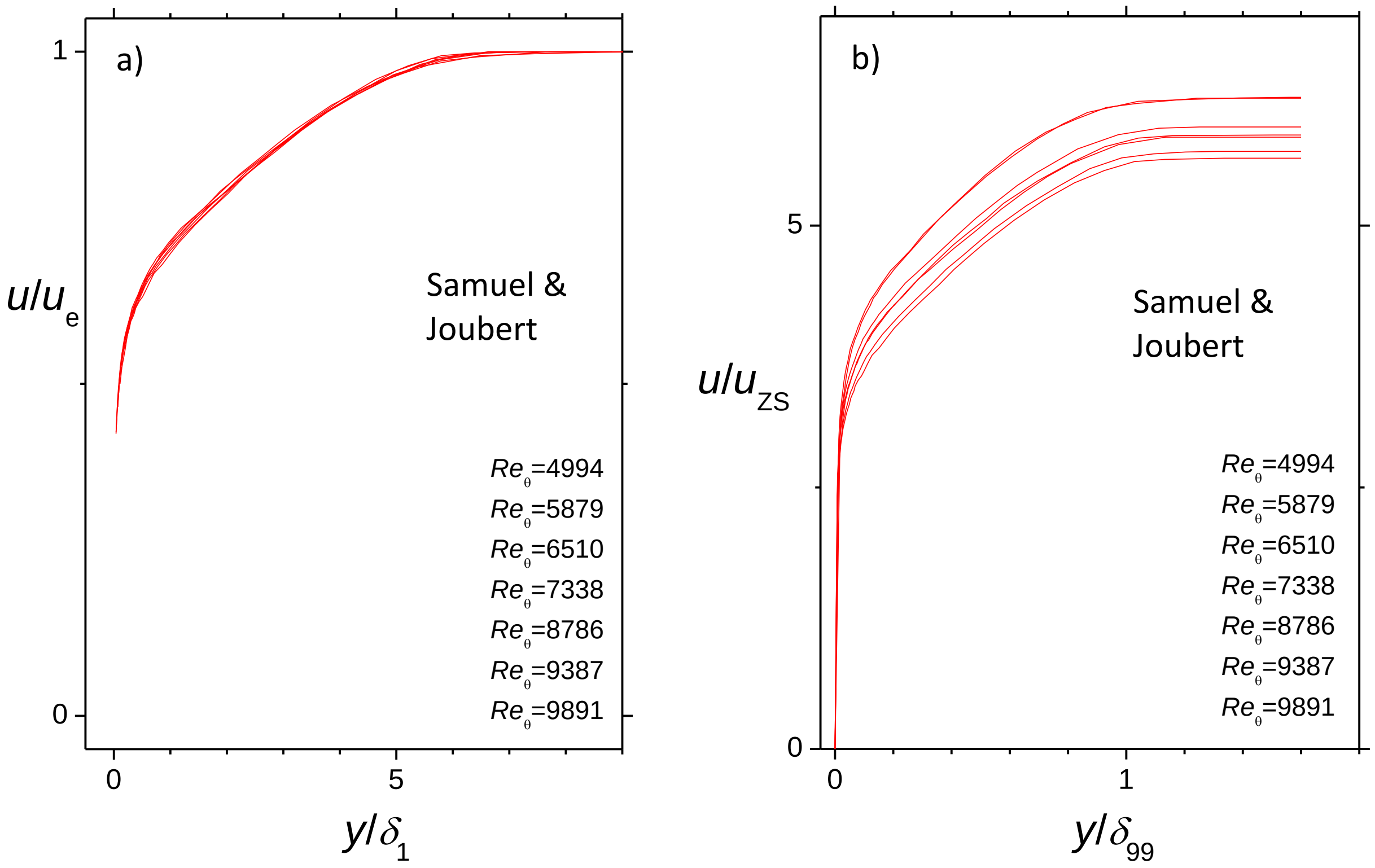

Fig. 6: Samuel and Joubert's [22] profile data plotted a) using $u_s = u_e$, b) using $u_s = u_{ZS}$, and c) using $u_s = u_\tau$. The $Re_\theta$ list identifies which profiles are plotted.

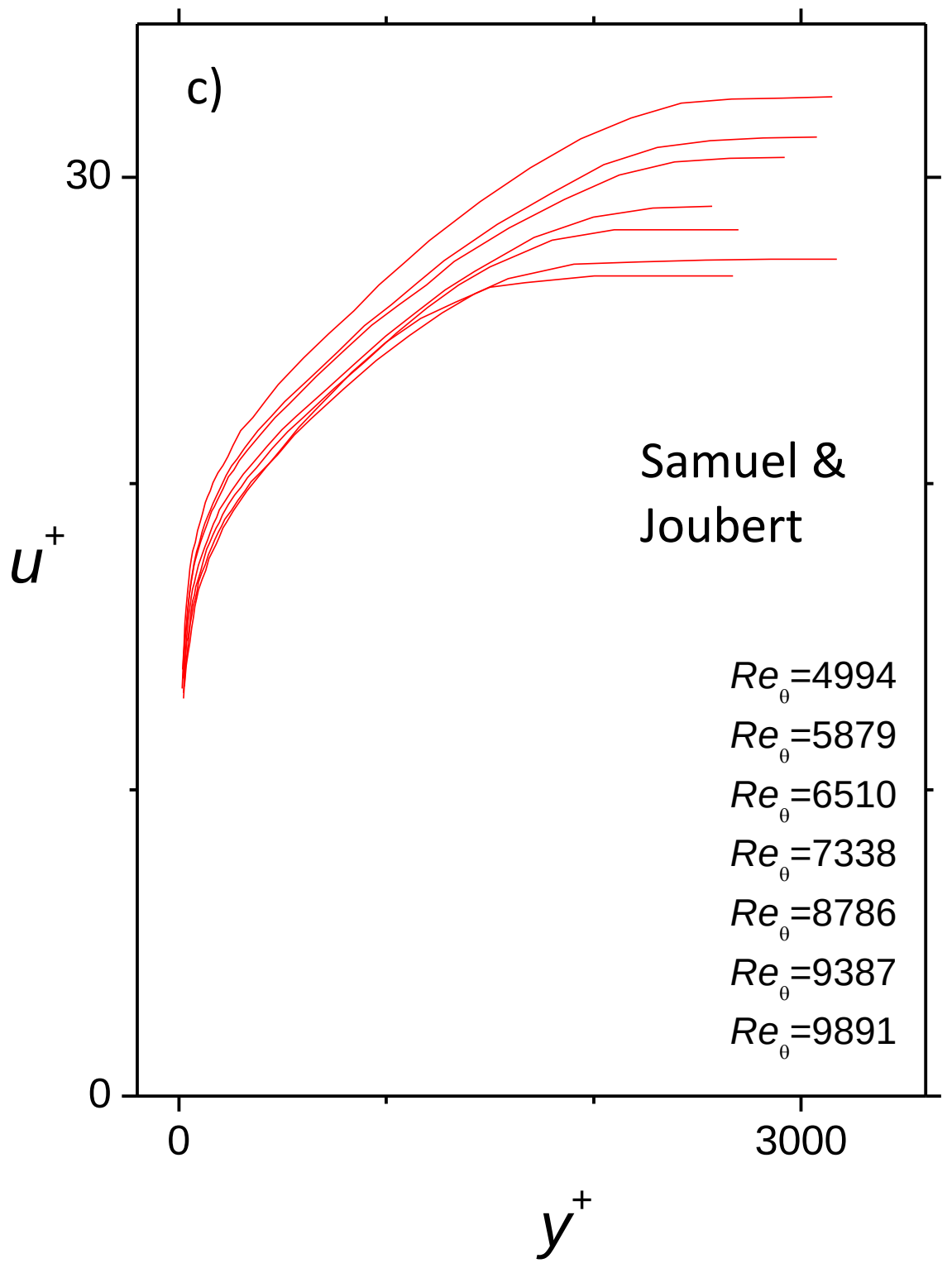

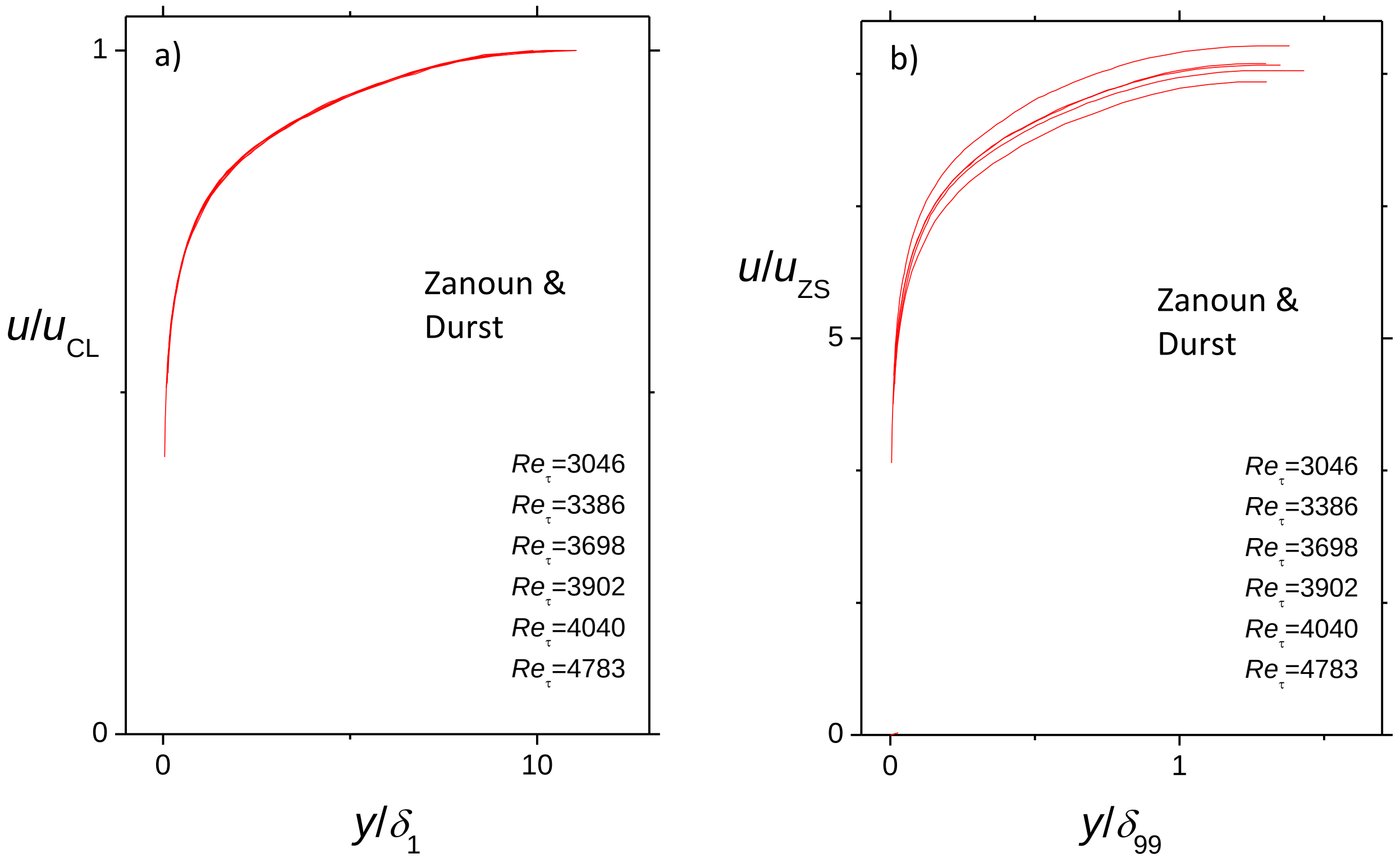

Fig. 7: Zanoun and Durst's [19] channel profile data plotted a) using $u_s = u_{CL}$, b) using $u_s = u_{ZS}$, and c) using $u_s = u_\tau$. The $Re_\theta$ list identifies which profiles are plotted.

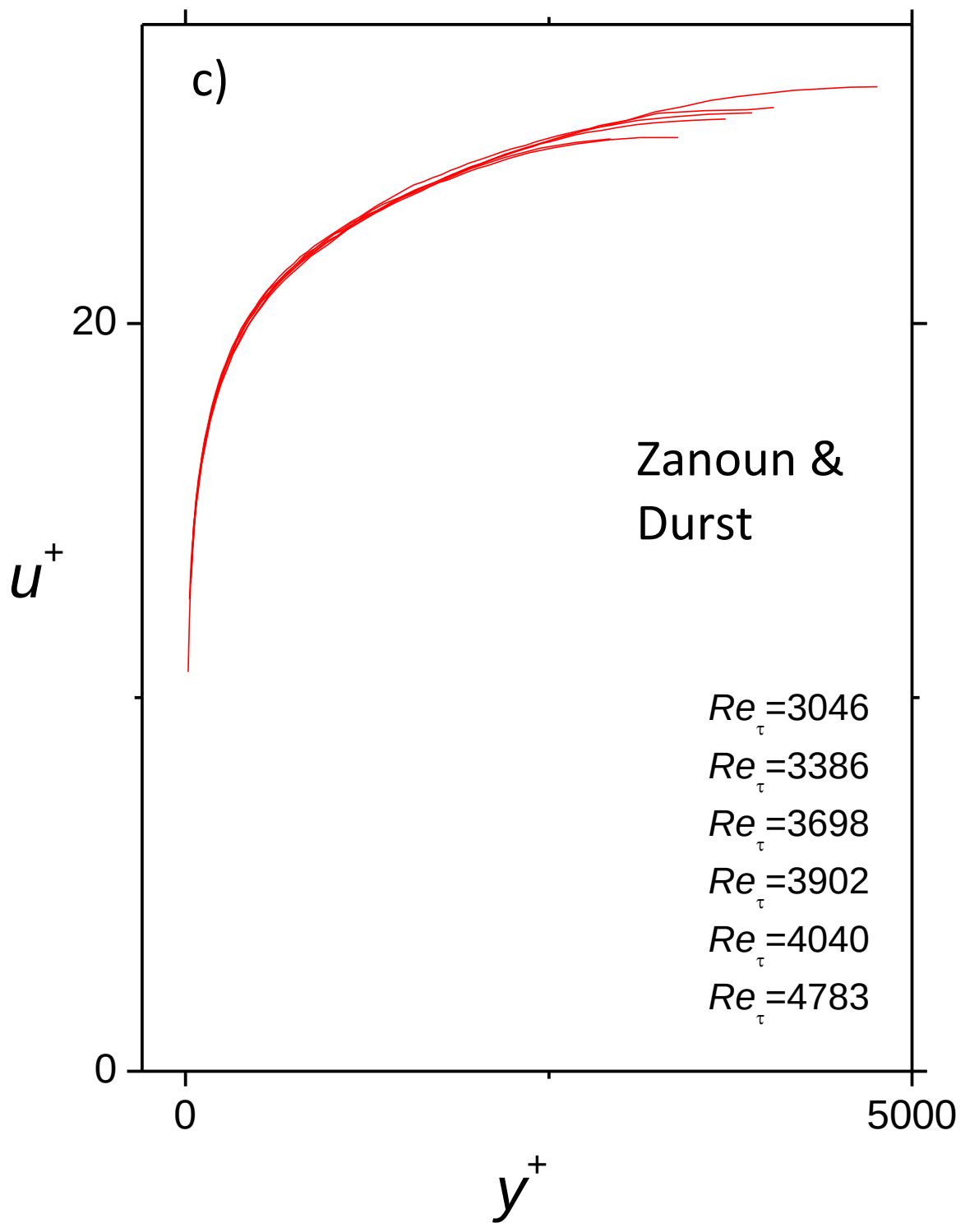

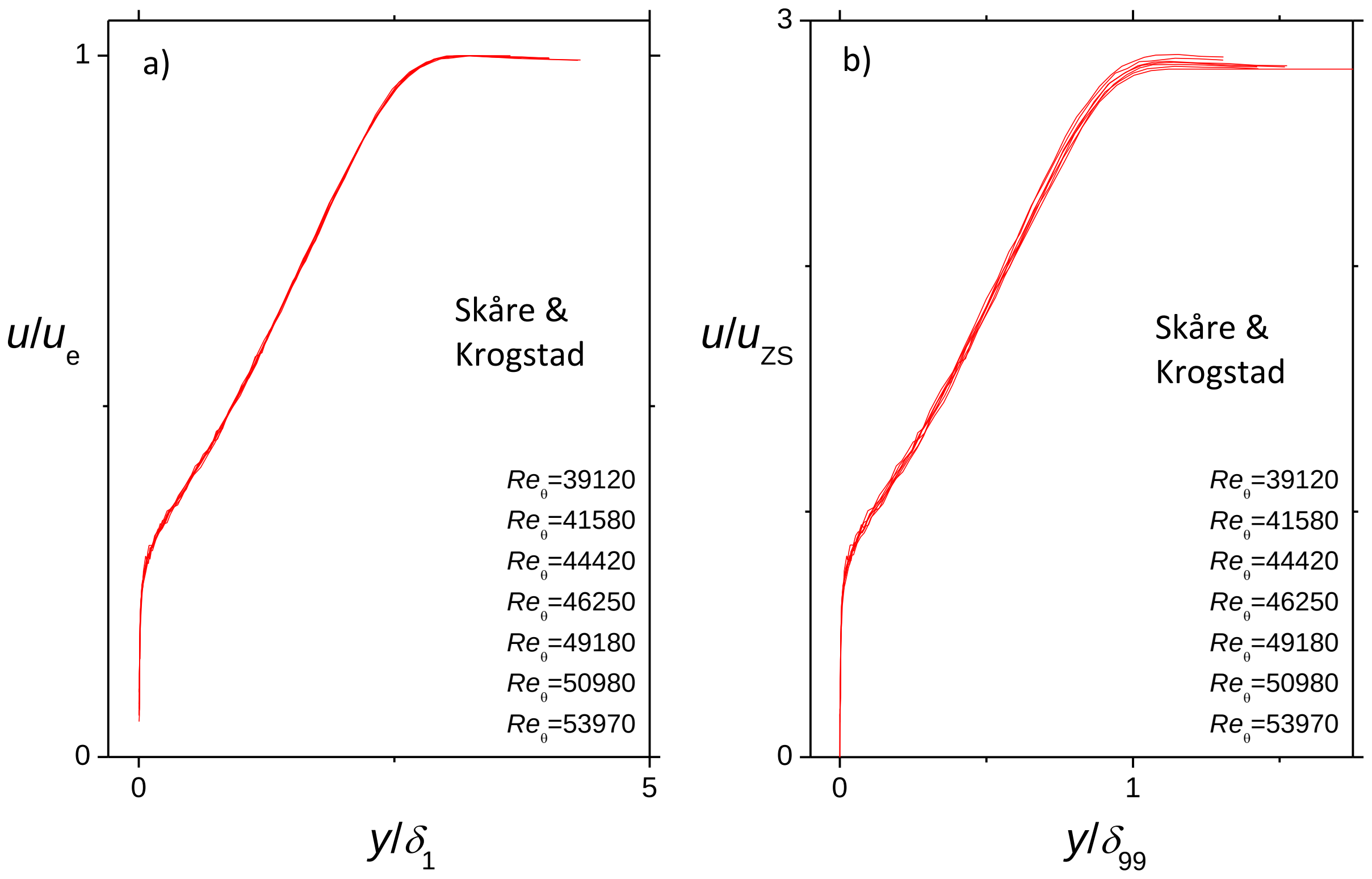

Fig. 8: Skåre and Krogstad's [5] profile data plotted a) using $u_s = u_e$, b) using $u_s = u_{ZS}$, and c) using $u_s = u_\tau$. The $Re_\theta$ list identifies which profiles are plotted.

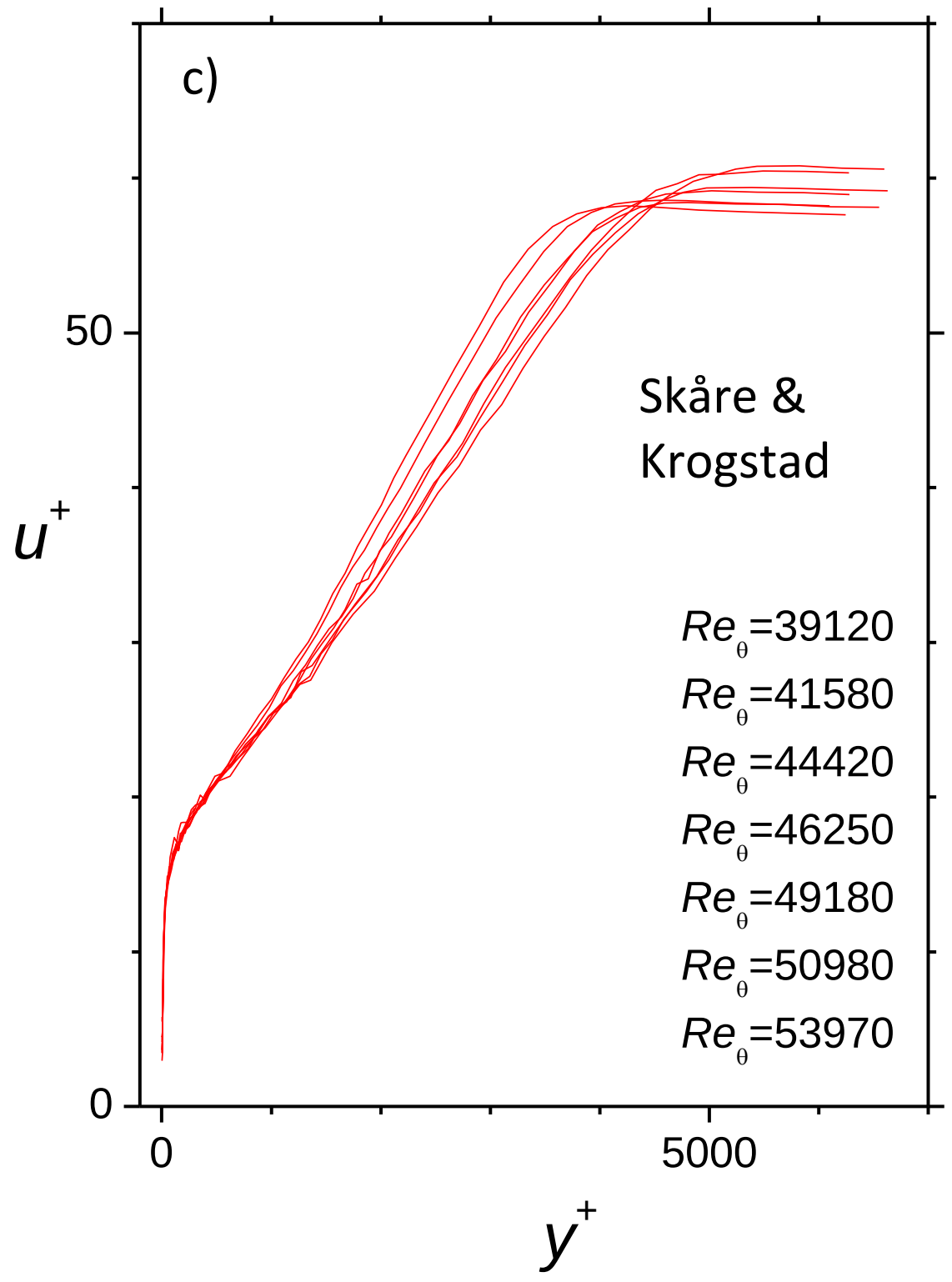

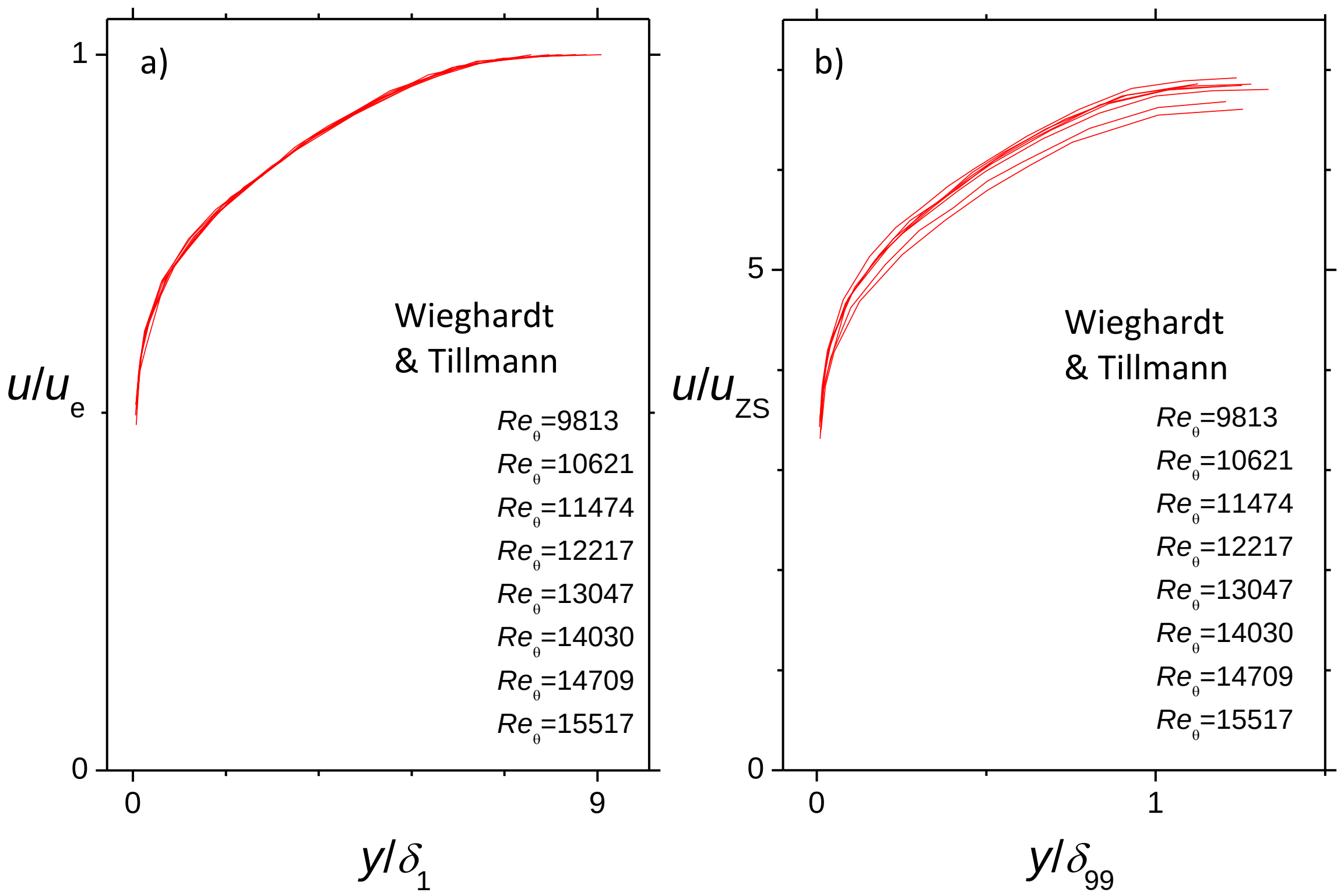

Fig. 9: Wieghardt and Tillmann's [23] profile data plotted a) using $u_s = u_e$, b) using $u_s = u_{ZS}$, and c) using $u_s = u_\tau$. The $Re_\theta$ list identifies which profiles are plotted.

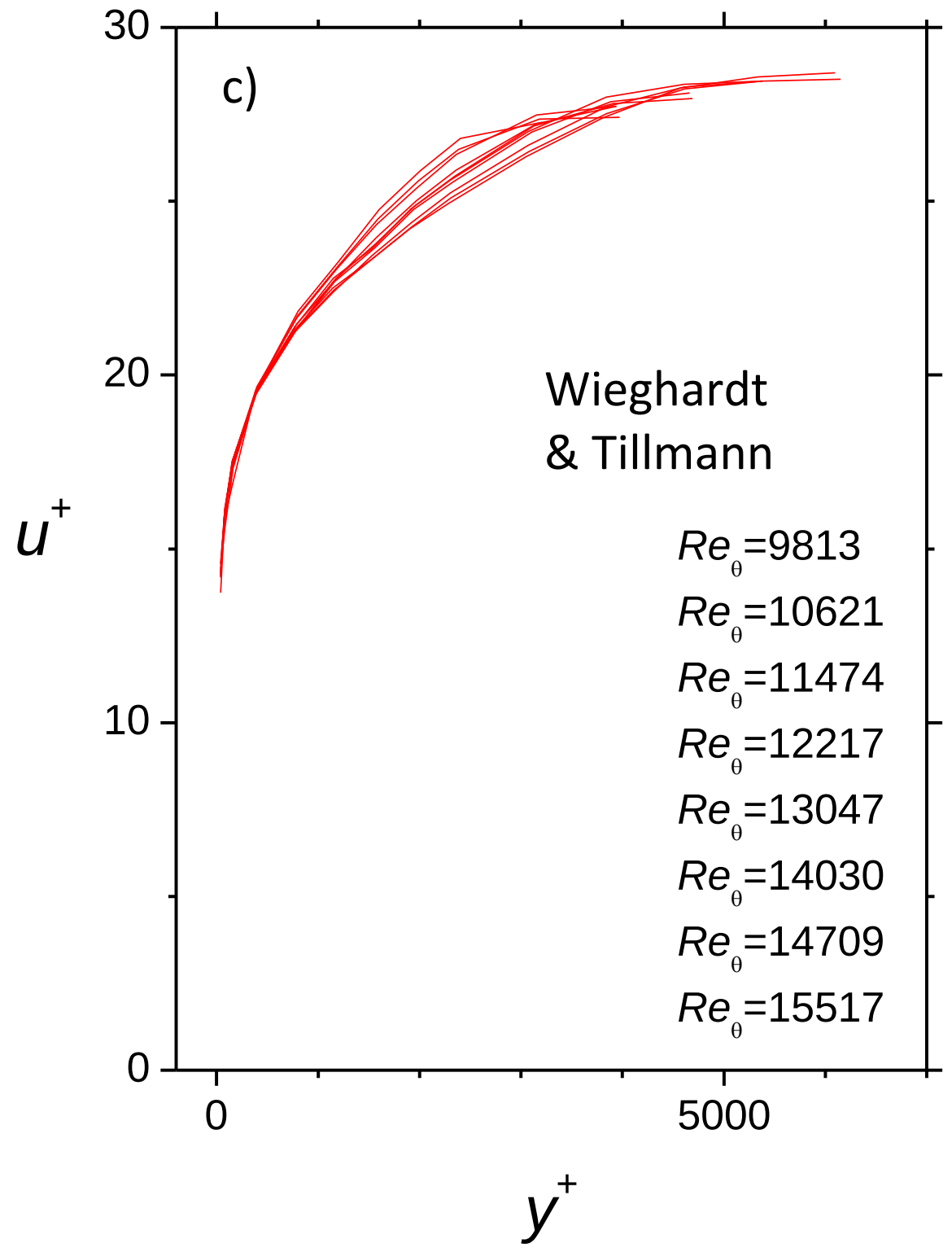

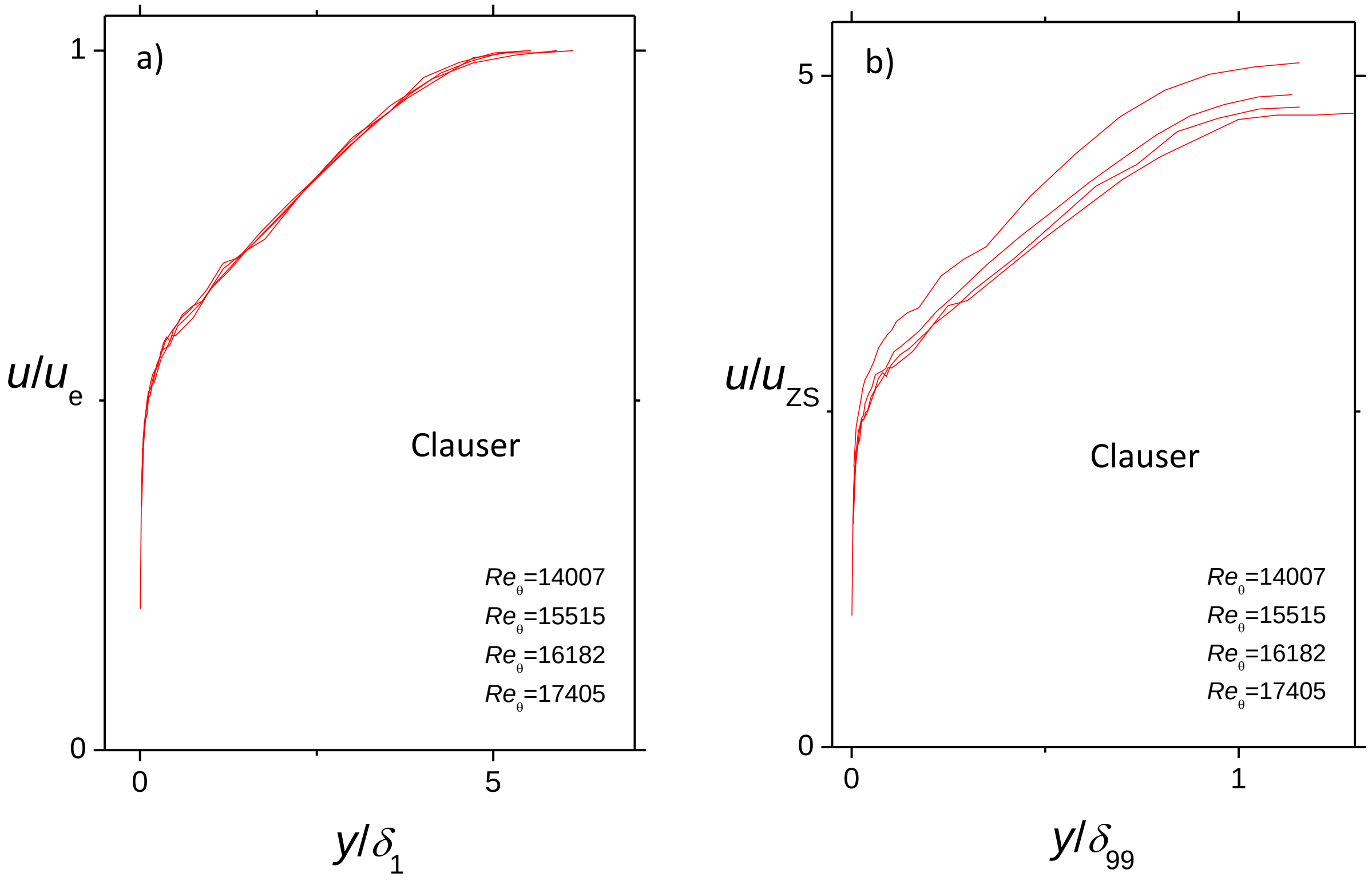

Fig. 10: Clauser's [15] profile data plotted a) using $u_s = u_e$, b) using $u_s = u_{ZS}$, and c) using $u_s = u_\tau$. The $Re_\theta$ list identifies which profiles are plotted.

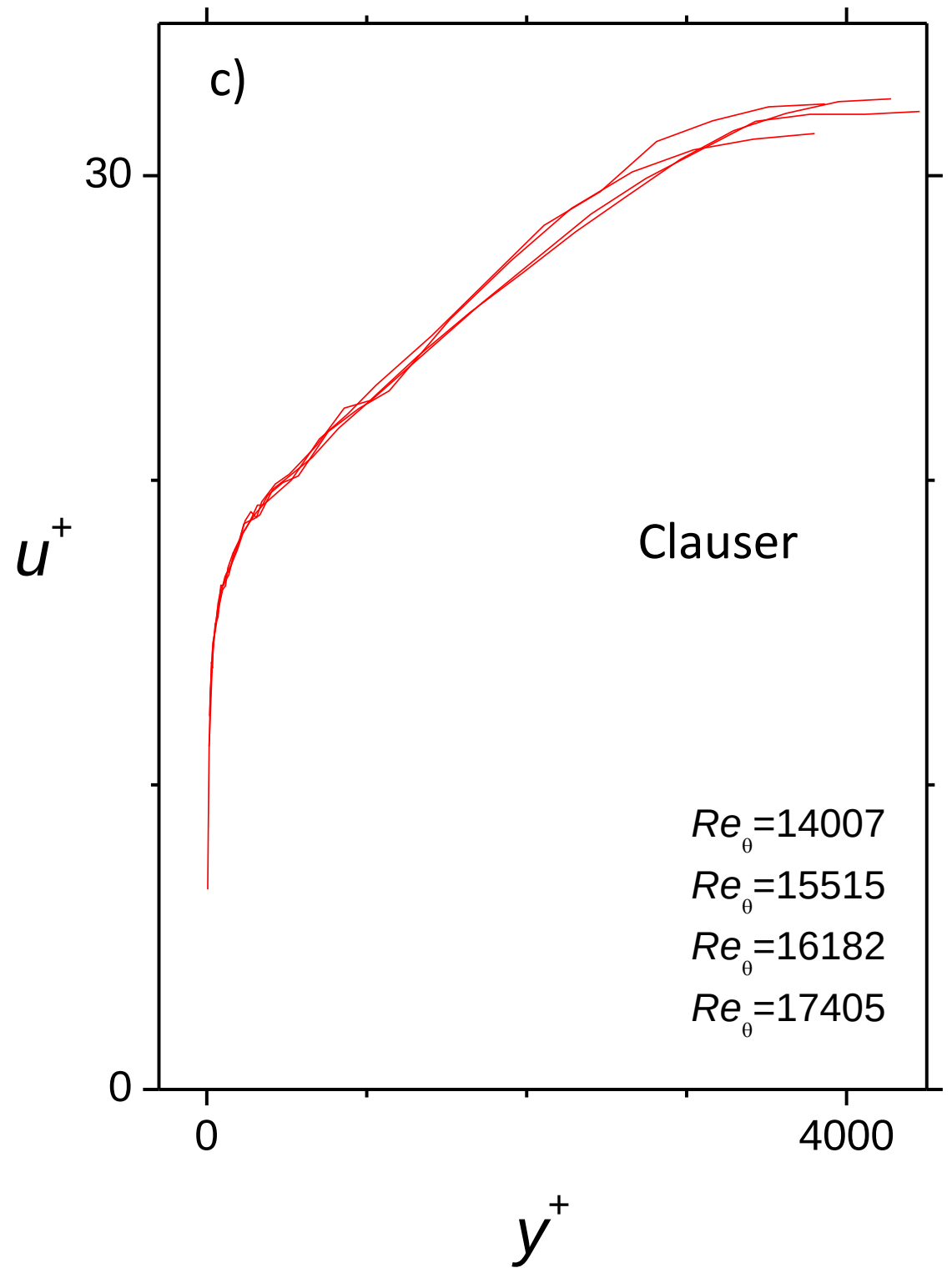

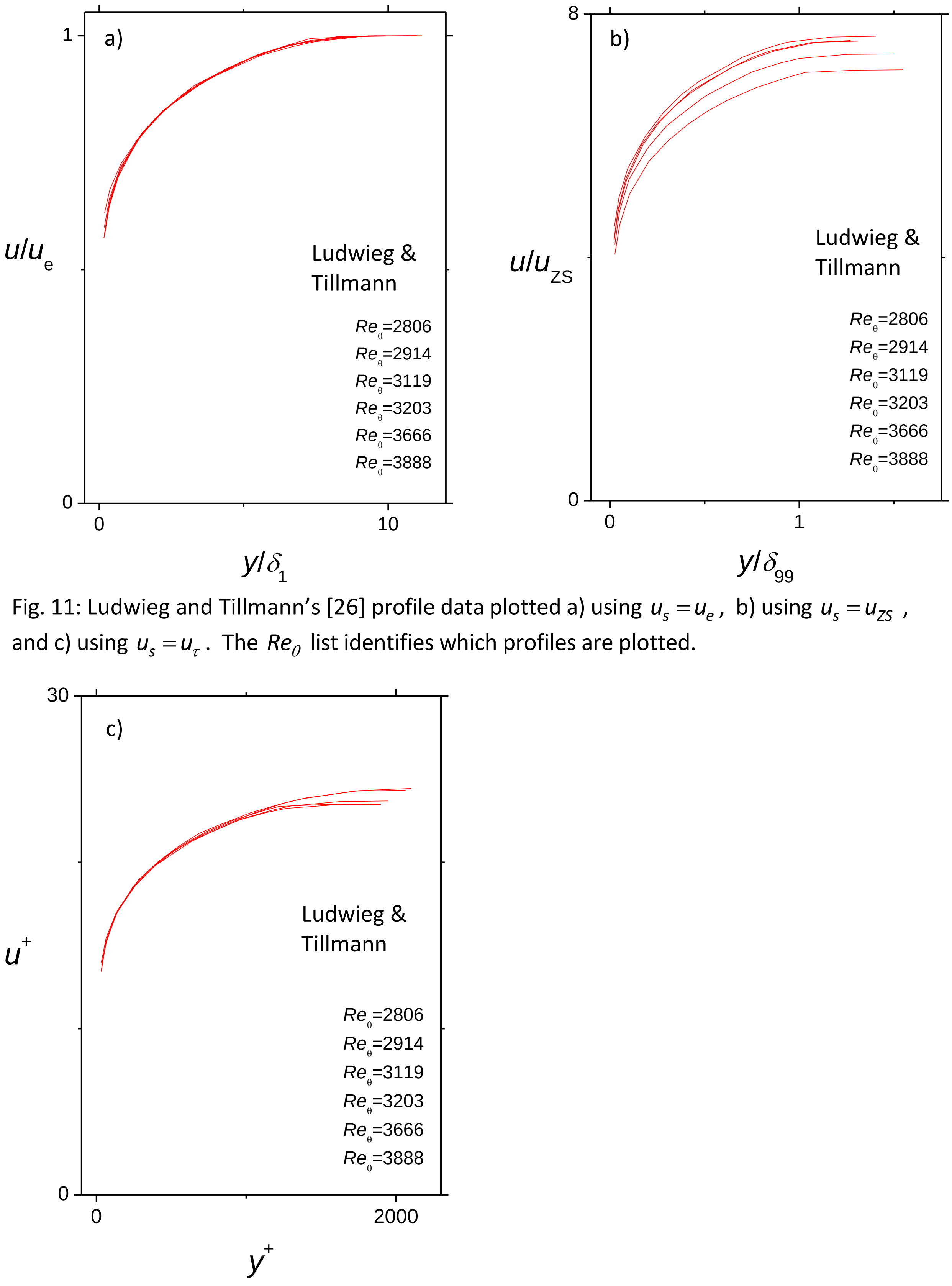

Fig. 11: Ludwieg and Tillmann's [26] profile data plotted a) using $u_s = u_e$, b) using $u_s = u_{ZS}$, and c) using $u_s = u_\tau$. The $Re_\theta$ list identifies which profiles are plotted.

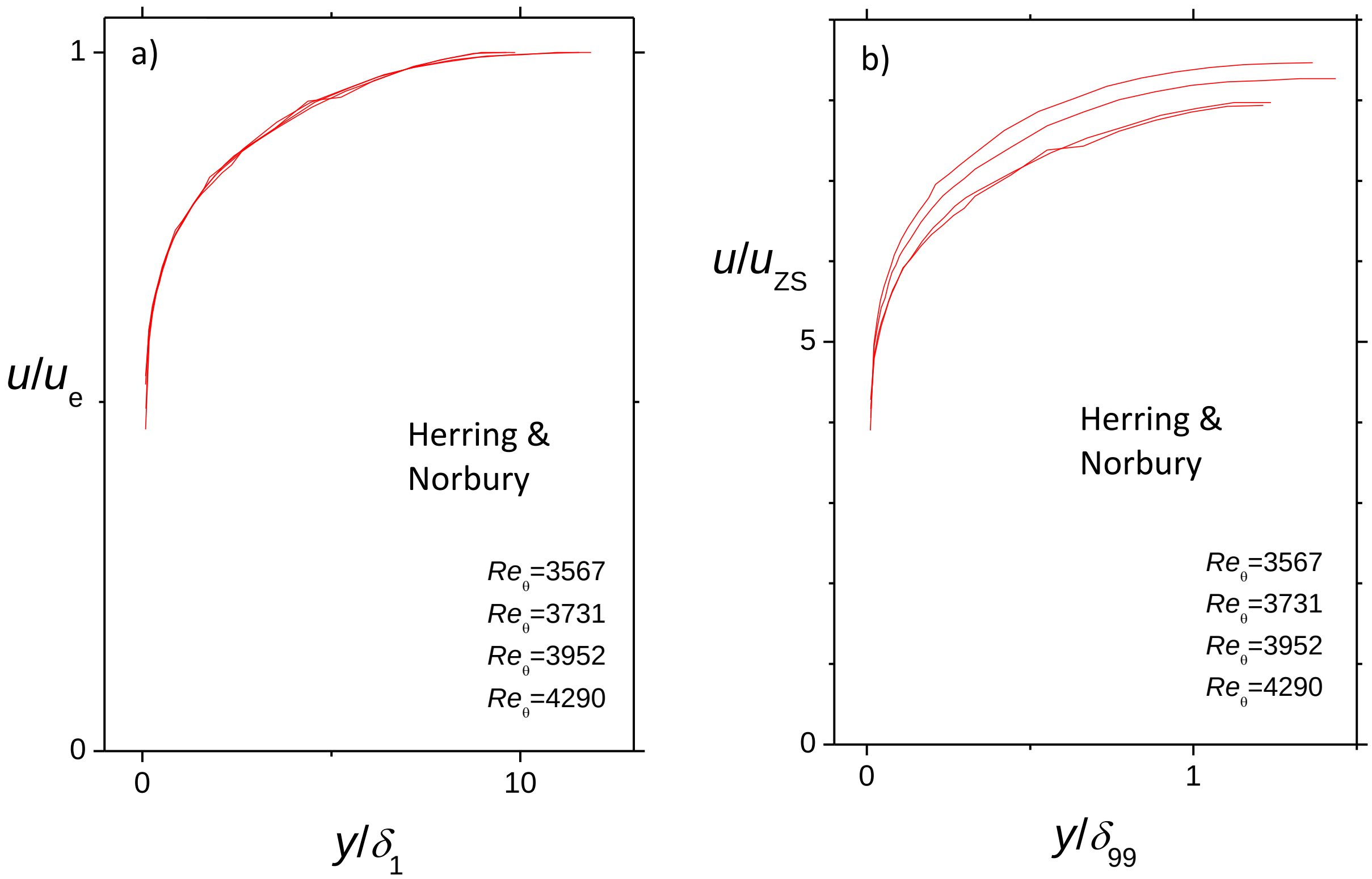

Fig. 12: Herring and Norbury's [20] data plotted a) using $u_s = u_e$, b) using $u_s = u_{ZS}$, and c) using $u_s = u_\tau$. The $Re_\theta$ list identifies which profiles are plotted.

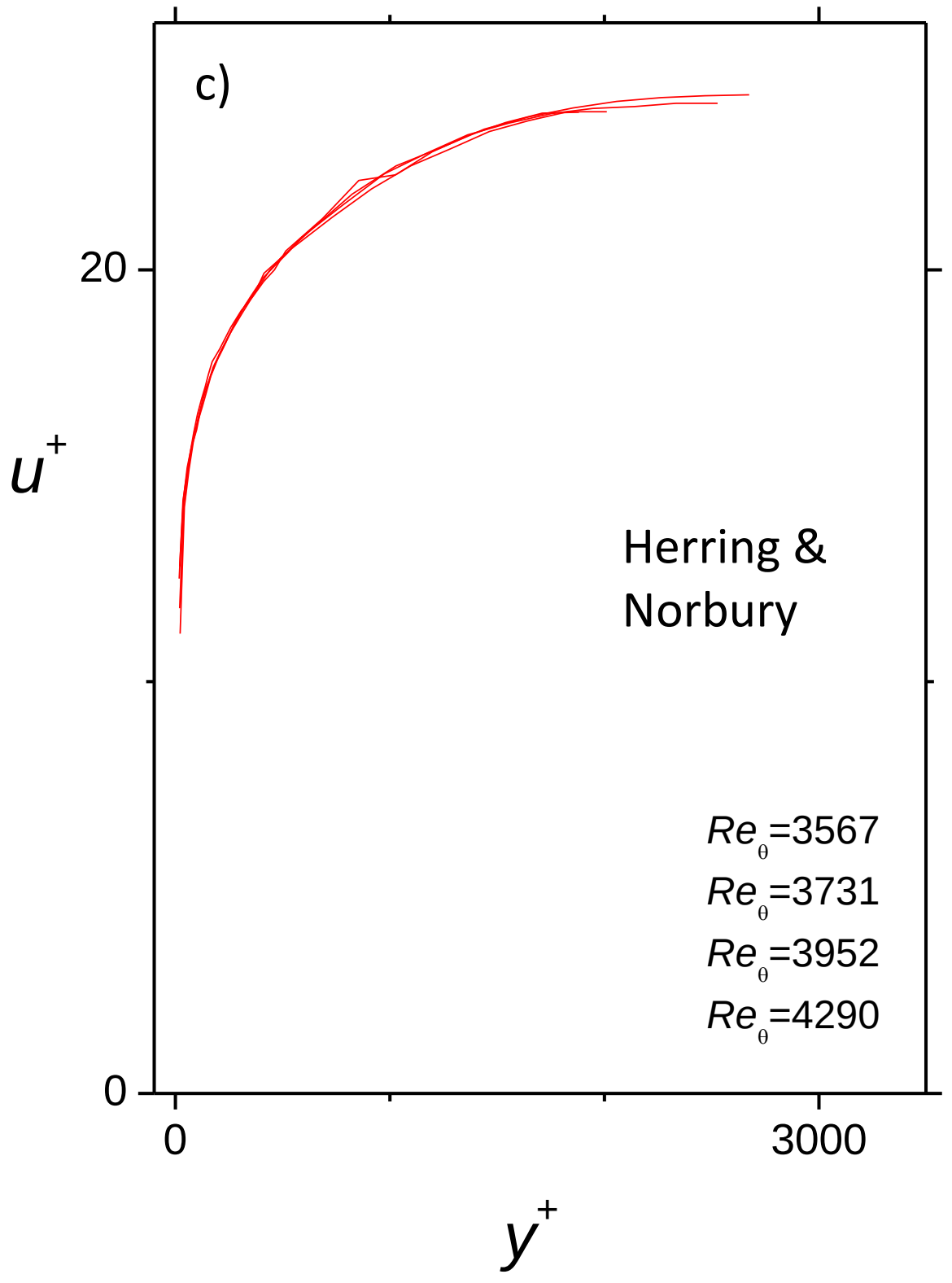

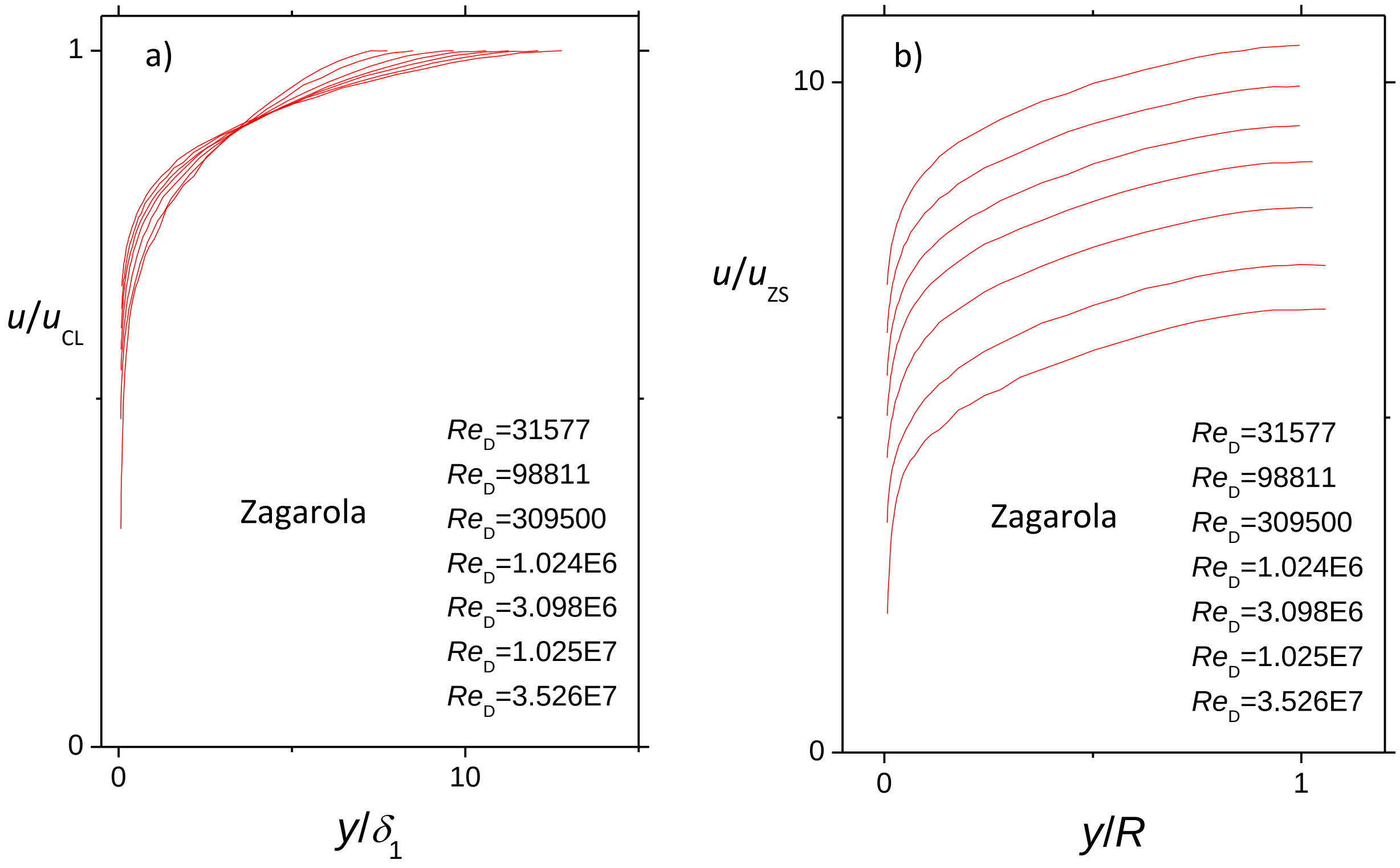

Fig. 13: Zagarola [11] profile data plotted a) using $u_s = u_{CL}$, b) using $u_s = u_{ZS} = u_{CL} - \bar{u}$, c) using $u_s = u_\tau$, and d) using $u_s = u_{ZS}$. The $Re_D$ list identifies which profiles are plotted.

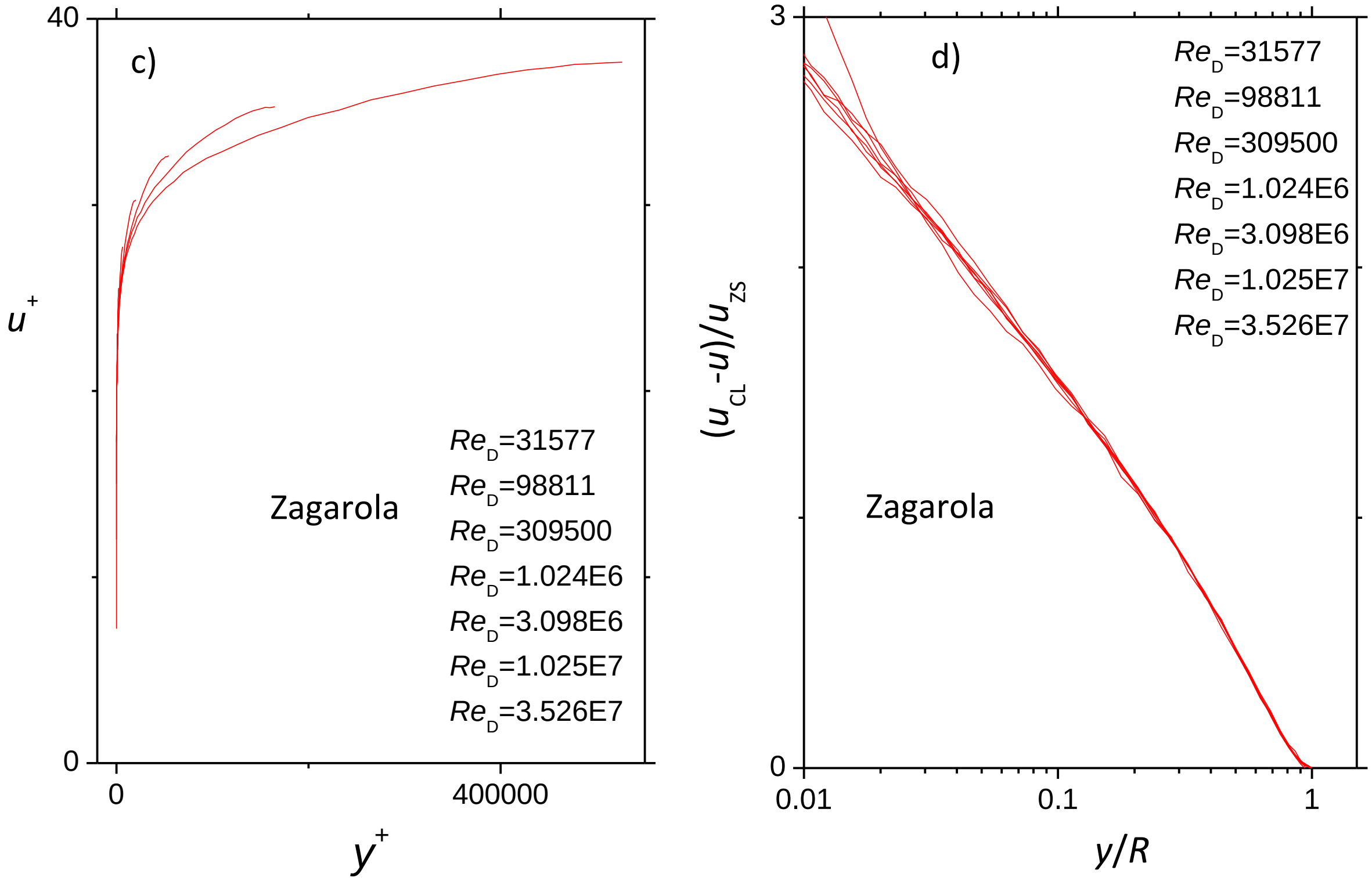

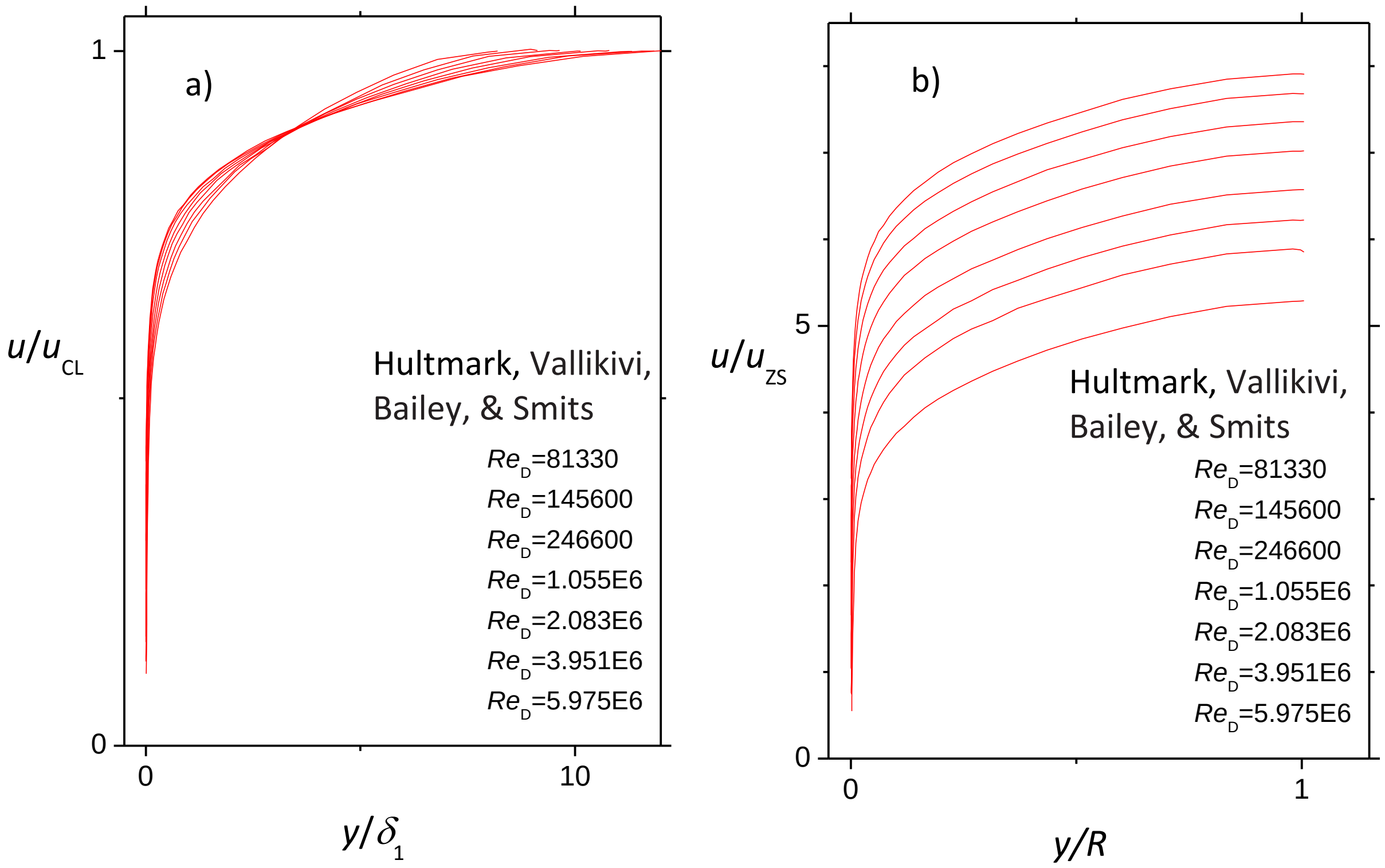

Fig. 14: Hultmark, Vallikivi, Bailey, and Smits [29] profile data plotted a) using $u_s = u_{CL}$, b) using $u_s = u_{ZS}$, c) using $u_s = u_\tau$, and d) using $u_s = u_{ZS}$. The $Re_D$ identifies which profiles are plotted.

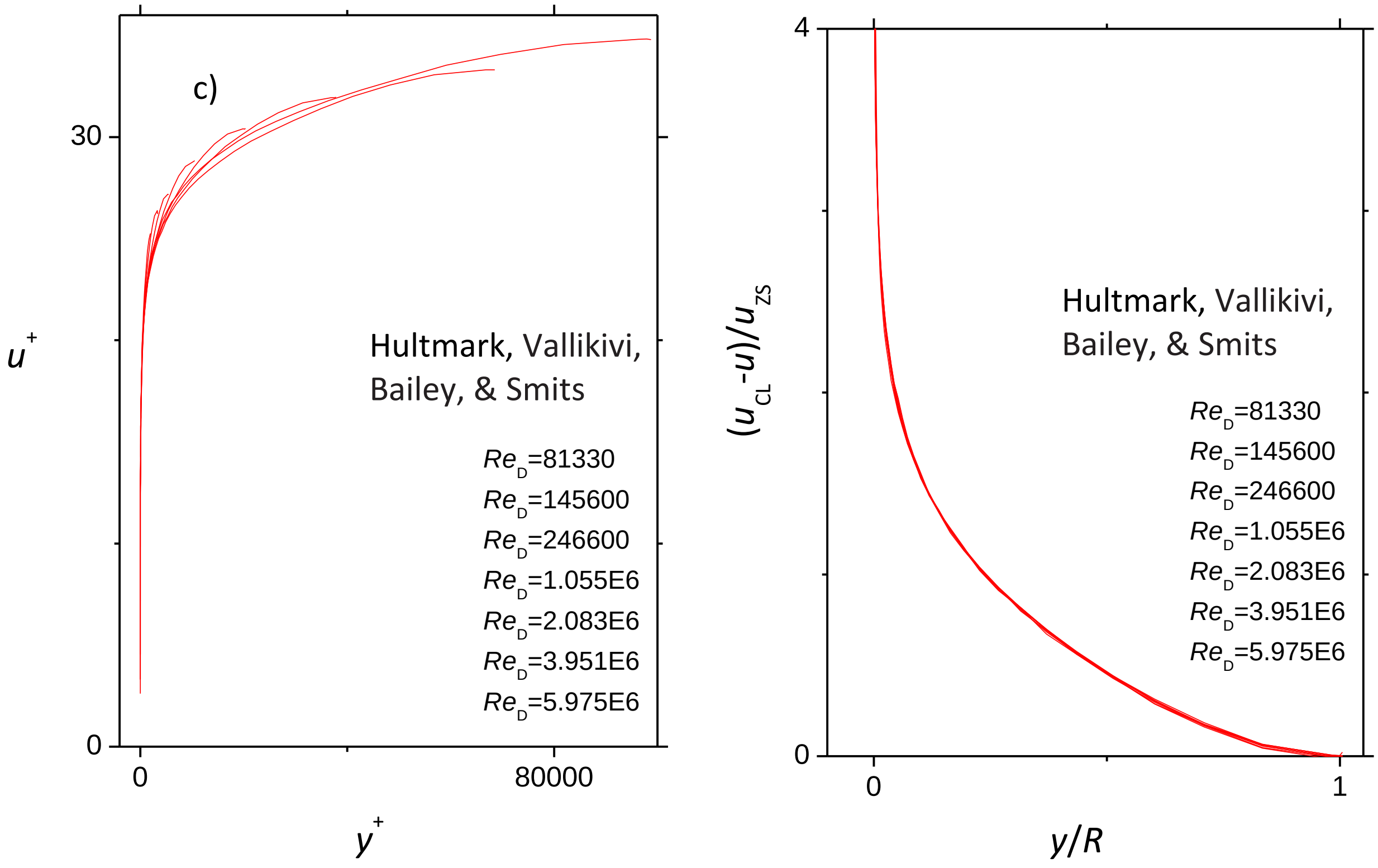

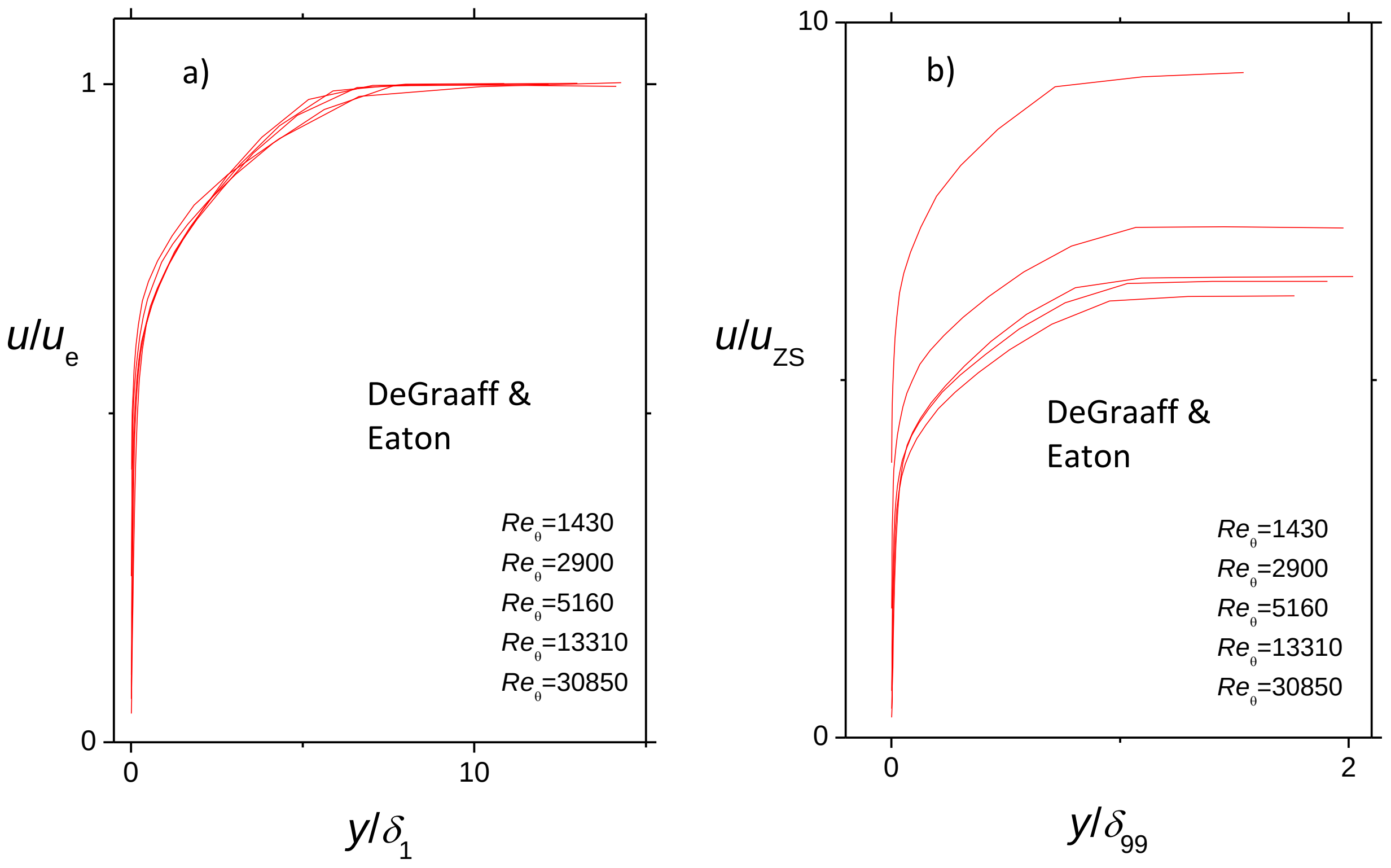

Fig. 15: DeGraaff and Eaton [28] data plotted a) using $u_s = u_e$, b) using $u_s = u_{zs}$, c) using $u_s = u_\tau$, and d) using $\delta_s = \delta_{99}$ and $u_s = u_e$. The $Re_\theta$ identifies which profiles are plotted.

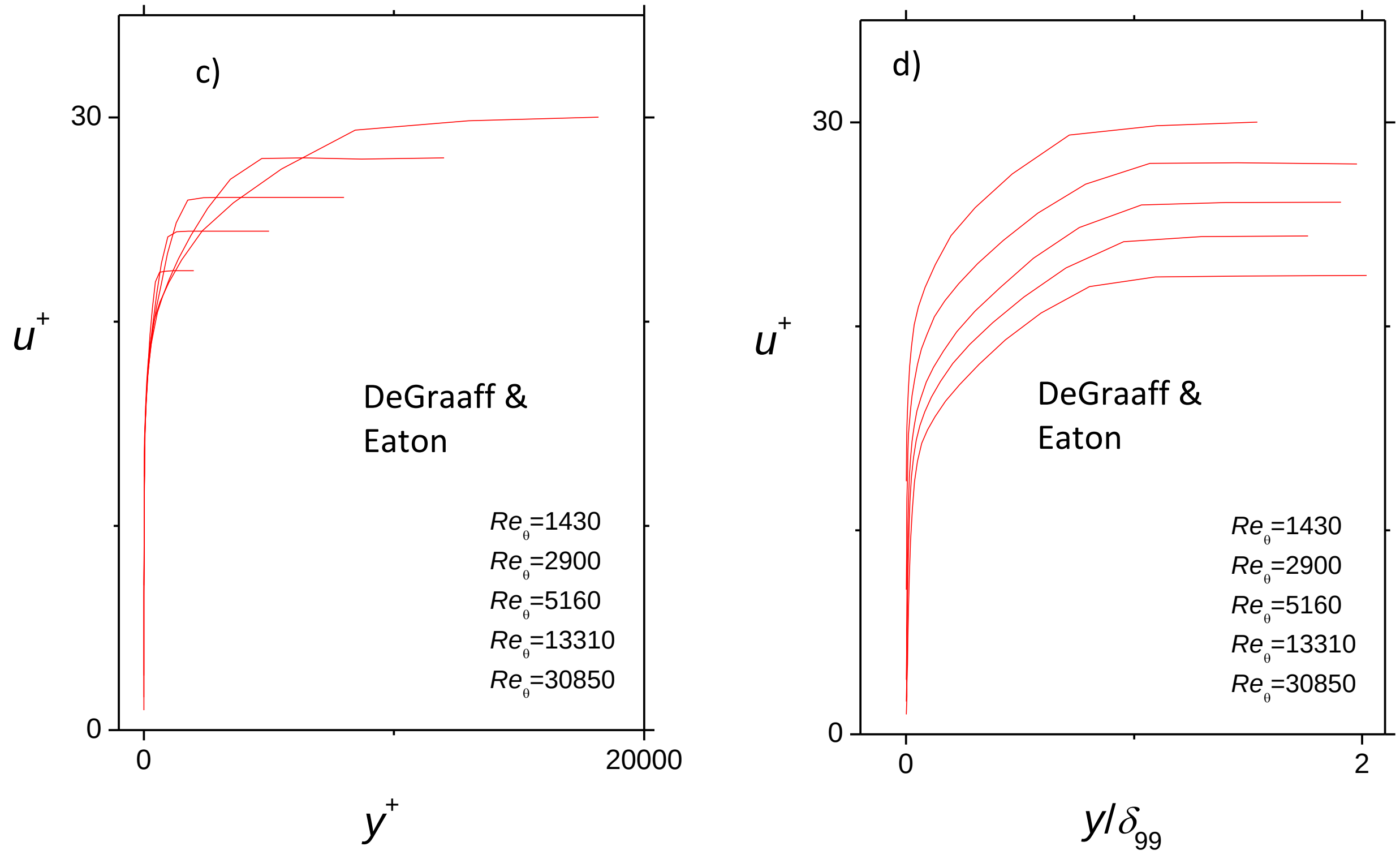

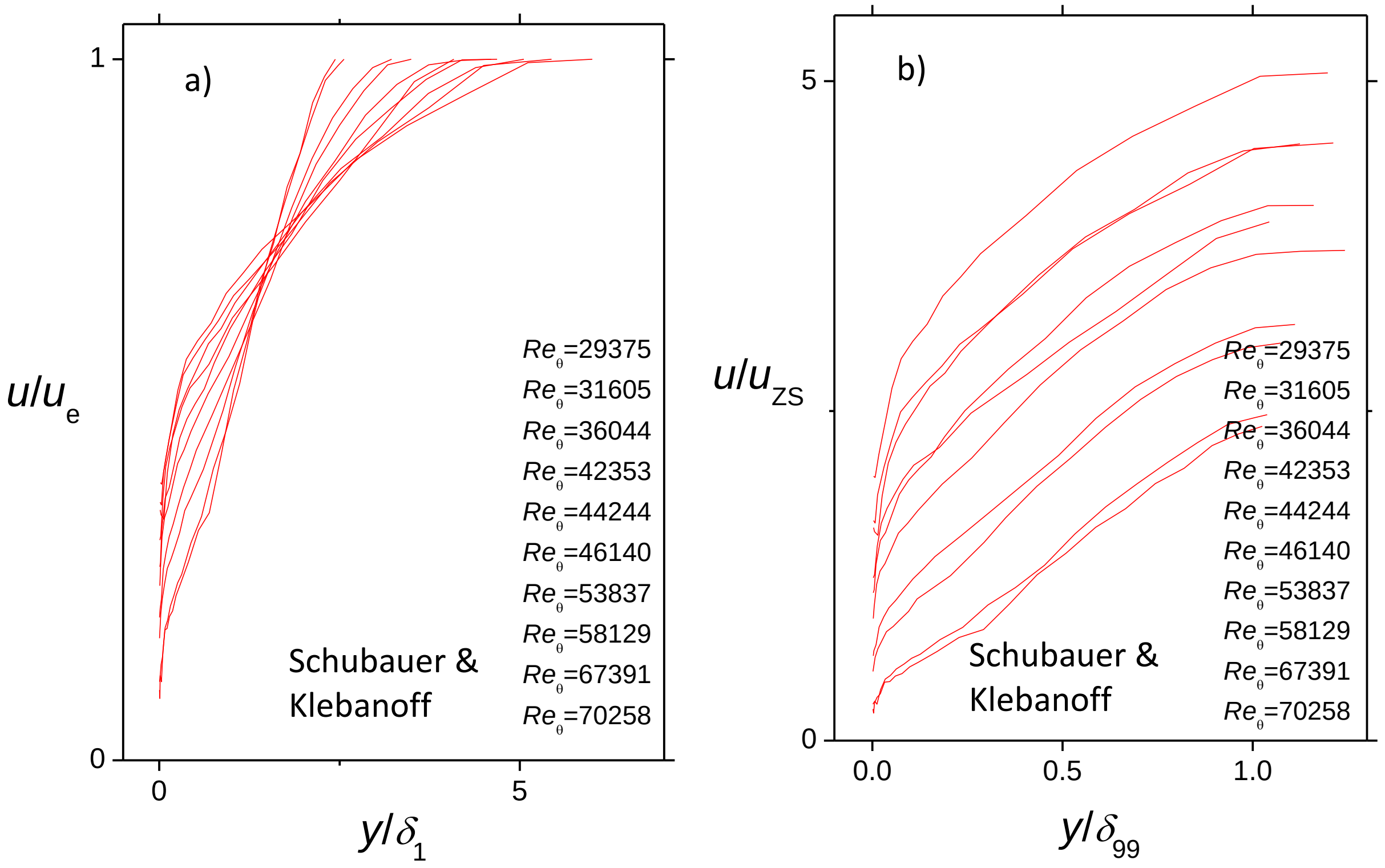

Fig. 16: Schubauer and Klebanoff [24] profile data plotted a) using $u_s = u_e$, b) using $u_s = u_{ZS}$, and c) using $u_s = u_\tau$. The $Re_\theta$ list identifies the plotted profiles.

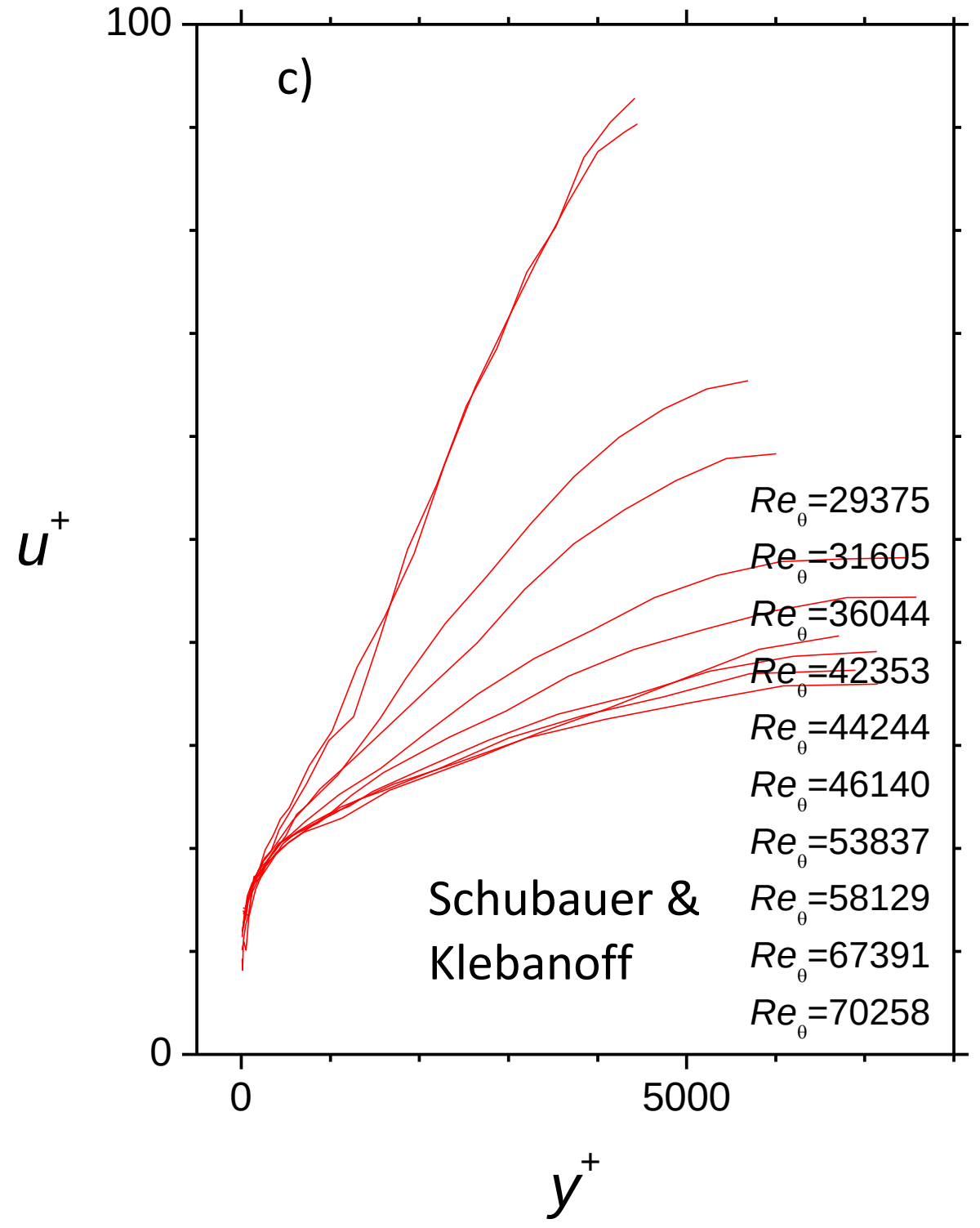